\documentclass[sigconf, nonacm]{acmart}
\usepackage{booktabs}
\usepackage{geometry}
\usepackage{subcaption}
\usepackage{capt-of}
\AtBeginDocument{%
  }

\setcopyright{none}
\copyrightyear{2026}
\acmYear{2026}
\acmDOI{XXXXXXX.XXXXXXX}
\acmConference[SIGIR '26]{Proceedings of the ACM Web Conference 2026}{July 20--24, 2026}{Melbourne, Australia}
\acmISBN{}

\begin{document}

\title{Sequences as Nodes for Contrastive Multimodal Graph Recommendation}

\author{Bucher Sahyouni}
\affiliation{%
  \institution{University of Surrey}
  \city{Guildford}
  \country{United Kingdom}
}
\email{bs00826@surrey.ac.uk}

\author{Matthew Vowels}
\affiliation{%
  \institution{The Sense, CHUV}
  \city{Lausanne}
  \country{Switzerland}
}
\affiliation{%
  \institution{Kivira Health}
  \city{New York}
  \state{NY}
  \country{USA}
}
\email{matthew.vowels@unil.ch}

\author{Liqun Chen}
\affiliation{%
  \institution{University of Surrey}
  \city{Guildford}
  \country{United Kingdom}
}
\email{liqun.chen@surrey.ac.uk}

\author{Simon Hadfield}
\affiliation{%
  \institution{University of Surrey}
  \city{Guildford}
  \country{United Kingdom}
}
\email{s.hadfield@surrey.ac.uk}

\renewcommand{\shortauthors}{Sahyouni et al.}

\begin{abstract}
To tackle cold start and data sparsity issues in recommender systems, numerous multimodal, sequential, and contrastive techniques have been proposed. While these augmentations can boost recommendation performance, they tend to add noise and disrupt useful semantics. To address this, we propose MuSICRec (Multimodal Sequence–Item Contrastive Recommender), a multi-view graph-based recommender that combines collaborative, sequential, and multimodal signals. We build a sequence-item (SI) view by attention pooling over the user's interacted items to form sequence nodes. We propagate over the SI graph, obtaining a second view organically as an alternative to artificial data augmentation, while simultaneously injecting sequential context signals. Additionally, to mitigate modality noise and align the multimodal information, the contribution of text and visual features is modulated according to an ID-guided gate.

We evaluate under a strict leave-two-out split against a broad range of sequential, multimodal and contrastive baselines. On the Amazon Baby, Sports and Elec datasets, MuSICRec outperforms the state-of-the-art baselines across all model types. In particular, we observe the largest gains for short-history users, effectively mitigating the sparsity and cold-start challenges that are ubiquitous in recommender systems. Our code is available at \\https://anonymous.4open.science/r/MuSICRec-3CEE/ and will be made publicly available.

\begin{figure}[htbp]
\centering
\includegraphics[width=\columnwidth]{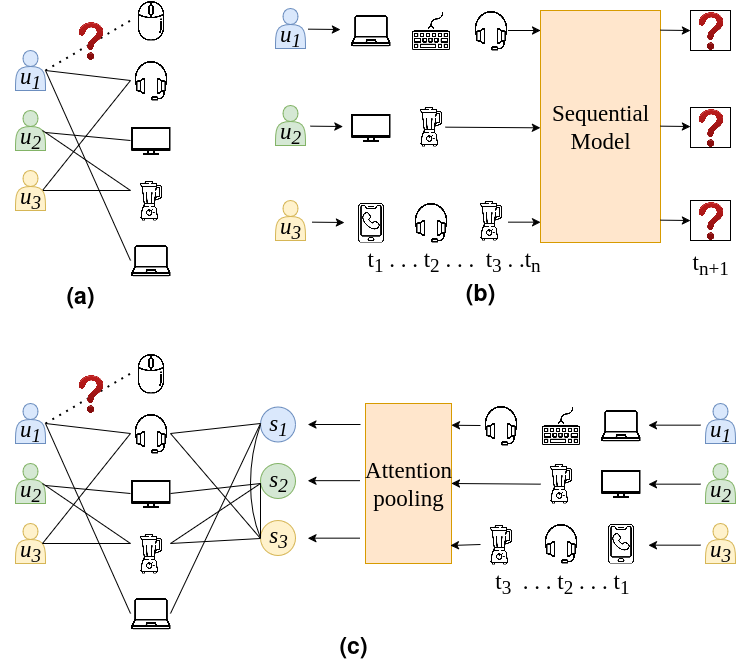}
\caption {Simplified diagram illustrating the difference between (a) collaborative filtering, (b) sequential methods and (c) MuSICRec. Collaborative filtering predicts user's preference for an item based on other users with similar preferences. Sequential recommenders predict next item a user would be interested in given their history. MuSICRec extracts sequence representations based on user history and treats them as nodes/entities to compliment collaborative filtering.}
\label{fig1}
\end{figure}
\end{abstract}

\ccsdesc[500]{Information systems~Recommender systems}
\ccsdesc[300]{Computing methodologies~Neural networks}
\ccsdesc[500]{Information systems~Multimedia information systems}

\keywords{Sequence-Item View, Multimodal, Sequential, Recommender System, GNN}

\received{20 February 2007}
\received[revised]{12 March 2009}
\received[accepted]{5 June 2009}

\maketitle

\section{Introduction}


Recommendation datasets are typically very sparse, necessitating methods that enable sharing of training signal between samples, such as collaborative filtering over user-item graphs, contrastive learning via alternative augmented views and extracting multimodal item-item relationships. Sequential recommendation is particularly valuable at capturing temporal user behaviour, but suffers more so from the same data sparsity issues. Figure \ref{fig1} presents a simplified illustration of collaborative filtering, sequential recommendation methods and how MuSICRec links the two approaches.
Existing graph-based recommender systems suffer from two limitations. \textbf{Firstly}, where contrastive learning (CL) is utilised to address label sparsity with self-supervised learning, CL approaches synthesise alternative views via heuristic perturbation of the user-item graph (e.g., node/edge dropout, random walks, clustering) \cite{Wu_2021}. Such augmentations may distort semantics and drop useful information, which could exacerbate the sparsity issue of inactive users. The success of such contrastive learning schemes is highly dependent on the view generator, affecting these models' generality. LightGCL \cite{cai2023lightgclsimpleeffectivegraph} argues this directly and utilises SVD to generate an alternative user-item view to preserve important user-item interaction semantics. Sequential CL models \cite{cl4srec} similarly carry out sequence augmentations, which could lead to the same issues.
\textbf{Secondly}, multimodal fusion is often not calibrated. While visual and textual data provide useful cues, there maybe misalignment between the modalites and with the preference/behaviour signal. This could lead to preference-irrelevant noise propagating into learned user and item representations. Recent analyses explicitly report such modality noise diffusion and amplification\cite{zhang2021mining, zhou2023tale, guo2024lgmreclocalglobalgraph}.Meanwhile, work that unifies sequential dynamics with multimodal information remains limited. Classic multimodal models and their denoising successors largely ignore sequences-as-structure, while sequential ones typically omit modalities. 

Beyond treating sequences as lists to be encoded, we treat every user's interaction history as a new class of node within a sequence-item (SI) graph, introducing sequence-to-sequence and sequence-to-item connections.
This choice is motivated by the evidence from sequential models showing sequential signals are highly informative for preference modelling \cite{kang2018self, sun2019, Qiu_2022,  du2023frequencyenhancedhybridattention}. Next-item prediction is greatly boosted by focusing on the most relevant past items through self-attention and bidirectional encoders (e.g., SASRec\cite{kang2018self}, BERT4Rec\cite{sun2019}). Treating sequences as nodes can expose structural patterns hidden by flat encoders. By passing attention-pooled sequence embeddings into the SI graph and propagating along sequence-item edges, MuSICRec derives an alternative view organically from the data's contextual behavioural topology rather than via synthetic data augmentation. This enables clean entity-level contrast (user \(u\) <-> sequence \(s_u\)) as the SI view complements the user-item (UI) view. At the same time, the SI graph captures a different signal from the UI bipartite graph. The UI graph encodes a global collaborative signal from user-item interactions, while an SI graph exposes intra-sequence structure (short-range and higher-order item transitions, co-occurrence order, and sequence/context signals) by treating each sequence as a node and connecting it to its constituent items and other sequences. This is akin to what sequential models exploit, but here it is made structural (sequence-as-nodes), so message passing can reveal patterns that UI alone can not. Practically, forming sequence embeddings via additive attention provides a stable, length-aware representation that emphasises salient items before message passing, while SI propagation exposes the inter-sequence co-ocurrence structure. This proves to be especially beneficial for users with short histories, effectively mitigating sparsity issues, as will later be empirically demonstrated.

Building on the insight that modality graphs can help with training sparsity, yet also inject noise if fused naively, MuSICRec anchors fusion in the ID (collaborative) space before any propagation. Instead of learning full modality graphs, we adopt a static multimodal item-item (II) graph, following FREEDOM \cite{zhou2023tale} (which showed that freezing the multimodal II graph can lead to higher stability and performance) and seed each item with its ID embedding plus an ID-conditioned gate that decides the proportion of visual and textual signal to inject. This mitigates cross-modality misalignment, respects dataset-specific salience of different modalities and limits modality noise from dominating collaborative signals, while letting multimodal content help where it is preference relevant, avoiding the coupling issues in MMGCN \cite{wei2019mmgcn} and the structure-learning overhead in LATTICE \cite{zhang2021mining}.
The MuSICRec model seeds the item state with the ID vector plus an ID-gated mixture of normalised visual/text projections, it then performs LightGCN-style propagation on the frozen II graph \cite{he2020lightgcn}.

In summary, this paper makes four main contributions:
\begin{itemize}
    \item Because collaborative filtering methods miss sequential/context signals, we introduce sequences as nodes, built by attention pooling over user history, within a SI graph.
    \item As artificially augmented views could disrupt behaviour semantics by dropping useful information, we propose using the SI view as an organic alternative view and carry out entity-level cross-view contrast between each user and their own sequence. 
    \item ID-guided cross-modal calibration is proposed to mitigate cross-modal misalignment, by gating the proportion of text and image features that are injected before propagation over the item-item graph.
    \item We propose a unified benchmarking procedure and perform an extremely broad evaluation, with baselines covering collaborative, sequential, multimodal and contrastive learning recommenders. We report comparative results on several Amazon datasets, as well as ablations, sensitivity to key hyperparameters, and an evaluation of different history lengths.
\end{itemize}

\section{Background}
\subsection{Multimodal Recommender Systems}
Multimodal recommender systems typically adopt graph-based or other deep learning methods to enrich collaborative filtering with visual and textual information. Earlier works, like VBPR \cite{he2016vbpr}, expand on classical BPR \cite{rendle2012bpr} by concatenating image embeddings to item-ID embeddings, while later models leverage attention mechanisms to discern user-specific modality preferences.

The ability of GNNs to capture higher-order relationships led to the introduction of several key approaches. MMGCN \cite{wei2019mmgcn} builds user-item graphs per modality, and the learned representations from each graph are aggregated to produce the final user and item embeddings. GRCN \cite{yu2021graph} improves on this by using a graph refining layer to down-weight noisy edges prior to message passing and aggregation. DualGNN \cite{wang2021dualgnn} takes a slightly different approach, building user-user co-occurrence graphs from modality-specific graphs via spectral clustering. This implicitly captures users' shifting preferences for different modalities and item-item relations. 

Mixing modality features into the user-item behaviour encoding process can lead to the amplification of modality noise and can harm the robustness of learned user and item embeddings. To this end, LATTICE \cite{zhang2021mining} and FREEDOM \cite{zhou2023tale} construct item-item graphs and separate modality propagation and aggregation from the main user-item graph. LATTICE \cite{zhang2021mining} builds item-item graphs per modality and jointly learns item embeddings and adjacency matrices. FREEDOM \cite{zhou2023tale} observed that freezing the LATTICE item-item graph leads to gains and is more lightweight, and therefore uses fixed item-item graphs, precomputed from raw multimodal features. Thus, GCN propagation depends only on structure. It also includes BPR losses to align user and modality embeddings and carries out degree-sensitive edge pruning on the user-item graph to denoise it. LGMRec \cite{guo2024lgmreclocalglobalgraph} similarly separates collaborative and modality signals, utilising a Local Graph Embedding module to learn separate interaction and multimodal user/item representations. It complements that with a Global Hypergraph Embedding module to capture global dependencies, also tackling sparsity.

Earlier non-multimodal contrastive learning models construct alternative views of the user-item graph with stochastic graph augmentations. SGL \cite{Wu_2021} uses edge/node dropout and random walks as a self-supervised signal learning. Later work demonstrates the unreliability of random perturbations with SimGCL \cite{simgcl} dropping graph augmentation in favour of random noise injection into the embeddings to form the contrastive view. LightGCL \cite{cai2023lightgclsimpleeffectivegraph} similarly avoids heuristic graph perturbations, generating views instead with SVD  spectral refinement, which simultaneously mitigates data sparsity and popularity bias as aligning with the global low-rank SVD view lowers reliance on high-degree (popular) nodes. BM3 \cite{zhou2023bootstrap} introduces contrastive learning to multimodal recommender systems via simple dropouts and three contrastive losses to optimise representations. 

\subsection{Sequential Recommender Systems}
Sequential methods optimise next-item prediction based on the user interaction history. With the popularisation of Deep learning, RNNs (particularly gated recurrent units) became popular in GRU4Rec \cite{hidasi2016}, but these were limited by sequential processing bottlenecks and vanishing gradient issues. Following the success of transformers in other fields through parallel processing and capturing long-range dependencies, SASRec \cite{kang2018self} exploited unidirectional self-attention to attend to the most relevant items in a user's history, while BERT4Rec \cite{sun2019} implemented a bidirectional Cloze objective to recover masked items from both left and right context.

Transformers face two limitations. Firstly, they lack strong inductive biases and therefore require more data to generalise well. The second is the `over-smoothing' issue, with self-attention acting like a low-pass filter that smoothes out high-frequency details in user behaviour \cite{shin2024attentive}. To counter these issues, recent methods utilise inductive bias or frequency-aware components. FEARec \cite{du2023frequencyenhancedhybridattention} performs self-attention on time and frequency domains with contrastive regularisation to capture low and high frequency signals. BSARec \cite{shin2024attentive} introduces a Fourier-based inductive bias that rebalances low and high frequencies to mitigate oversmoothing. On the other hand, FMLPRec \cite{fmlprec} reframes the problem as learnable signal filtering and removes self-attention, replacing it with stacked filter-enhanced blocks.

\section{Methodology}

\begin{figure*}[t]
\centering
\includegraphics[width=0.8\textwidth]{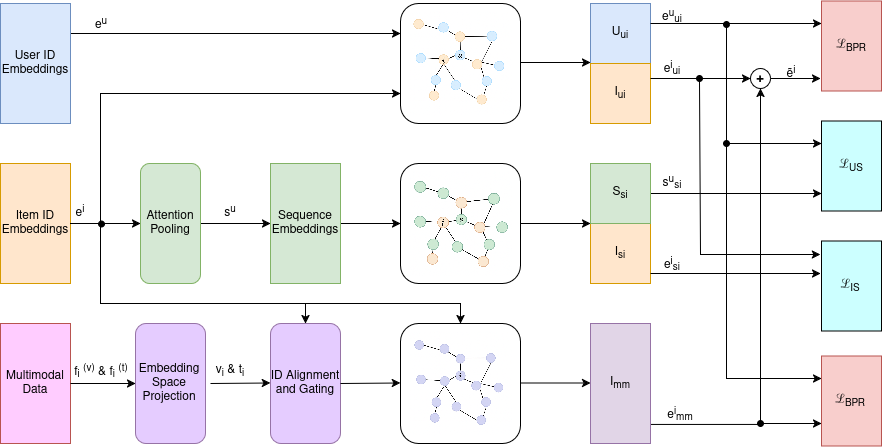}
\caption {MuSICRec Architecture}
\label{fig2}
\end{figure*}

\subsection{Problem Formulation and Notation}
This section will formalise the problem setup and the overall architecture of the proposed MuSICRec framework, as illustrated in Figure~\ref{fig2}.

\paragraph{Entities and data}
$\mathcal{U}=\{1,\dots,M\}$ and $\mathcal{I}=\{1,\dots,N\}$ denotes the set of users and items, respectively. A binary interaction matrix is defined, 
$\mathbf{R}\in\{0,1\}^{M\times N}$, such that $R_{u i}=1$ indicates that user $u\in\mathcal{U}$
interacted with item $i\in\mathcal{I}$ (implicit feedback). 
A time-ordered interaction history is observed for every user $u$:
\begin{equation}
\mathbf{s}_u=\big[(i_{u,1}, t_{u,1}),\ldots,(i_{u,T_u}, t_{u,T_u})\big],
\quad i_{u,k}\in\mathcal{I},\; t_{u,1}<\cdots<t_{u,T_u},
\end{equation}
where $T_u$ is the sequence length.

\paragraph{Base embeddings}
We store trainable ID-embedding tables for users and items, 
$\mathbf{U}\in\mathbb{R}^{M\times d}$ and $\mathbf{I}\in\mathbb{R}^{N\times d}$,
and denote by $\mathbf{e}^{u}\in\mathbb{R}^{d}$ and $\mathbf{e}^{i}\in\mathbb{R}^{d}$
the $u$-th and $i$-th rows of $\mathbf{U}$ and $\mathbf{I}$, respectively, and by
$d$ the embedding dimension. 

A set of sequence nodes $\mathcal{S}$ is introduced, where each user is associated with one sequence node $s_u\in\mathcal{S}$. Similar to user and item embeddings, we maintain a trainable sequence embeddings table:
\begin{equation}
\mathbf{S}\in\mathbb{R}^{|\mathcal{S}|\times d}.
\end{equation}

\paragraph{Multimodal item content}
As well as its learned embedding, each item $i$ also has pre-extracted visual and textual features, 
$\mathbf{f}^{(v)}_{i}\in\mathbb{R}^{d_v}$ and $\mathbf{f}^{(t)}_{i}\in\mathbb{R}^{d_t}$.
Linear projection matrices $\mathbf{W}_v\in\mathbb{R}^{d_v\times d}$ and
$\mathbf{W}_t\in\mathbb{R}^{d_t\times d}$ map these features into the shared space.

\paragraph{Graphs}
MuSICRec is designed to exploit collaborative filtering to improve sample efficiency. To that end, three graph structures are constructed to enable information sharing between the above entities.
(i) a user--item (UI) bipartite graph with adjacency $\mathbf{A}_{\mathrm{ui}}$,
(ii) a sequence--item (SI) bipartite graph with adjacency $\mathbf{A}_{\mathrm{si}}$ and
(iii) a multimodal item--item (MM) graph with adjacency $\mathbf{A}_{\mathrm{mm}}$. The exact construction of these adjacency matrices will be discussed in more detail in the following sections.


\subsection{User-Item Graph}
The simplest graph in MuSICRec is designed to allow information sharing between users who interacted with similar items, and items which were selected by similar users. It is formed directly from the interaction matrix $\mathbf{R}$  as a bipartite graph over the users and items.
\begin{equation}
\hat{\mathbf{A}}_{\mathrm{ui}}
=\begin{bmatrix}
\mathbf{0} & \mathbf{R}\\[2pt]
\mathbf{R}^{\!\top} & \mathbf{0}
\end{bmatrix}\!,
\qquad
\mathbf{A}_{\mathrm{ui}}
=\mathbf{D}^{-\frac{1}{2}}\hat{\mathbf{A}}_{\mathrm{ui}}\mathbf{D}^{-\frac{1}{2}},
\end{equation}
where $\mathbf{D}$ is the diagonal degree matrix of $\hat{\mathbf{A}}_{\mathrm{ui}}$.

To allow longer range connections, information is repeatedly propagated through the graph via $L_{\mathrm{ui}}$ sequential layers using simple neighbourhood averaging
\begin{equation}
\mathbf{Z}^{(l+1)} \;=\; \mathbf{A}_{\mathrm{ui}}\;\mathbf{Z}^{(l)}, \qquad l=0,\ldots,L_{\mathrm{ui}}-1.
\end{equation}
As in LightGCN \cite{he2020lightgcn}, no transforms, biases or nonlinearities are applied between layers.

For the first layer input, the user and item embedding are stacked 
\begin{equation}
\mathbf{Z}^{(0)}=\begin{bmatrix}\mathbf{U}\\ \mathbf{I}\end{bmatrix}\in\mathbb{R}^{(M+N)\times d},
\end{equation}
while the final $L_{\mathrm{ui}}{+}1$ layer outputs are aggregated by a uniform mean:
\begin{equation}
\bar{\mathbf{Z}} \;=\; \frac{1}{L_{\mathrm{ui}}+1}\sum_{l=0}^{L_{\mathrm{ui}}}\mathbf{Z}^{(l)},
\end{equation}
before finally splitting back out into user and item blocks,
\begin{equation}
\bar{\mathbf{Z}}=
\begin{bmatrix}
{\mathbf{U_{ui}}}\\ {\mathbf{I_{ui}}}
\end{bmatrix},\qquad
{\mathbf{U_{ui}}}\in\mathbb{R}^{M\times d},\;{\mathbf{I_{ui}}}\in\mathbb{R}^{N\times d}.
\end{equation}
We denote the embeddings produced by the UI-graph as 
$\mathbf{e}^{u}_{\mathrm{ui}}$ (the $u$-th row of ${\mathbf{U_{ui}}}$) and
$\mathbf{e}^{i}_{\mathrm{ui}}$ (the $i$-th row of ${\mathbf{I_{ui}}}$).

\subsection{Sequence-Item Graph}

To allow us to consider sequential relationships without our graphical model, each interaction sequence $s_u$ in the dataset is aggregated to a single embedding vector using additive attention pooling.
To achieve this for sequence $s$ with length $\ell_s$, let 
$\mathbf{H}_s=[\mathbf{e}^{i_1},\ldots,\mathbf{e}^{i_{\ell_s}}]\in\mathbb{R}^{\ell_s\times d}$ be the concatenation of the item embeddings. 
This is passed through a learned attention function to obtain the attention weights
\begin{equation}
\mathbf{a}_s = \mathrm{softmax}\!\left(\mathbf{e}_s\right),
\quad
\mathrm{where}
\quad
\mathbf{e}_s = \mathbf{v}_a^{\top}\tanh(\mathbf{W}_a \mathbf{H}_s^{\top})\in\mathbb{R}^{\ell_s}, 
\end{equation}
and $\mathbf{W}_a\in\mathbb{R}^{h\times d}$ and $\mathbf{v}_a\in\mathbb{R}^{h}$ are learned weights.

Finally, the item embedding sequence is combined with learned positional embeddings $\mathbf{P}_{1:\ell_s}\in\mathbb{R}^{\ell_s\times d}$ to form 
\begin{equation}
\mathbf{X}_s \;=\; \mathbf{H}_s + \mathbf{P}_{1:\ell_s},
\end{equation}
which is modulated with the attention weights through a weighted sum aggregation\footnote{other interchangeable attention variants such as sinusoidal pooling are possible and will be examined in Section \ref{ablation}}
\begin{equation}
\mathbf{s}^{(0)}_s \;=\; \sum_{t=1}^{\ell_s} a_{s,t}\,\mathbf{X}_{s,t}.
\end{equation}

For efficiency, we 
cache the matrix $\mathbf{S}$ containing these attention pooled sequence embeddings.
For every batch during training, we employ a refresh strategy where only the embeddings of the sequences contained in the batch are recomputed from the latest item-ID embedding table. 
Then at the start of each epoch, we refresh \emph{all} subsequence vectors in the cache. 

\paragraph{SS and SI edges.}
To integrate these sequence embeddings into the overall MuSICRec graphical structure, we construct a \emph{weighted} adjacency over $\mathcal{S}\cup\mathcal{I}$. This
combines both sequence--sequence (SS) and sequence--item (SI) links, enabling information sharing between the training signal of similar sequences, and between sequences and their constituent items.

SS edges are constructed and weighted using the Jaccard similarity between \emph{sets} of items to determine their similarity:
\begin{equation}
\textstyle \mathrm{jac}(s,s') \;=\; \frac{\lvert \mathrm{set}(\mathcal{I}_s)\cap \mathrm{set}(\mathcal{I}_{s'})\rvert}{\lvert \mathrm{set}(\mathcal{I}_s)\cup \mathrm{set}(\mathcal{I}_{s'})\rvert}.
\end{equation}
We construct the graph by adding an undirected SS edge $(s,s')$ if $\mathrm{jac}(s,s')\ge \tau$, with the edge weight stored as the Jaccard similarity, $w_{ss}=\mathrm{jac}(s,s')$.
%
%
We also add SI edges to connect $s$ to each constituent item $i\in\mathcal{I}_s$ with unit weight.

The resulting weighted adjacency is $\hat{\mathbf{A}}_{\mathrm{si}}\in\mathbb{R}^{(n_s+N)\times(n_s+N)}$, where $n_s=|\mathcal{S}|$.
We apply symmetric GCN normalization
\begin{equation}
\mathbf{A}_{\mathrm{si}} \;=\; \mathbf{D}^{-1/2}\,\hat{\mathbf{A}}_{\mathrm{si}}\,\mathbf{D}^{-1/2},
\quad \mathbf{D}=\mathrm{diag}\!\left(\hat{\mathbf{A}}_{\mathrm{si}}\mathbf{1}\right),
\end{equation}
and represent $\mathbf{A}_{\mathrm{si}}$ as a sparse tensor for efficient computation.

\paragraph{Propagation and outputs.}
As for the User--Item graph, we perform $L_{\mathrm{si}}$ steps of linear propagation on the Sequence--Item graph:
\begin{equation}
\mathbf{Z}^{(\ell+1)}_{\mathrm{si}} \;=\; \mathbf{A}_{\mathrm{si}}\,\mathbf{Z}^{(\ell)}_{\mathrm{si}},
\qquad \ell=0,\ldots,L_{\mathrm{si}}-1.
\end{equation}
The input at layer 0 is the stacked sequence and item embeddings:
\begin{equation}
\mathbf{Z}^{(0)}_{\mathrm{si}} \;=\; \begin{bmatrix}
\mathbf{S} \\[2pt] \mathbf{I}
\end{bmatrix} \in \mathbb{R}^{(N+n_s)\times d},
\end{equation}
and the final output is split back out into sequences and items
\begin{equation}
\mathbf{Z}^{(L_{\mathrm{si}})}_{\mathrm{si}}
\;=\; \begin{bmatrix}
\mathbf{S}_{\mathrm{si}} \\[2pt] \mathbf{I}_{\mathrm{si}}
\end{bmatrix},
\end{equation}
where $\mathbf{I}_{\mathrm{si}}\in\mathbb{R}^{N\times d}$ are the SI-updated item
representations and $\mathbf{S}_{\mathrm{si}}\in\mathbb{R}^{n_s\times d}$ are the updated
sequence representations.  These feed into the fusion stage together with the user–item
branch and the multimodal branch.

\subsection{Multimodal Item-Item Graph}
The third graph module within the MuSICRec framework is designed to allow the sharing of information between items that do not co-occur, but which are similar according to their supplementary multimodal data.
To this end, we build item--item graphs from the raw image and text data, then project these modality features into the shared ID embedding space. Each item is then \emph{seeded} with an ID-anchored, per-item \emph{gated} mix of modality embeddings before linear propagation. 
As suggested by \cite{zhou2023tale} we keep the connectivity of the II graph frozen during training to improve stability.
Learning item--item structures online \cite{zhang2021mining} is powerful but expensive and can drift; freezing the $k$NN content graph keeps training lean and reproducible.

\paragraph{Feature projections.}
Each item $i$ carries raw visual and textual features, $\mathbf{f}^{(v)}_i\!\in\!\mathbb{R}^{d_v}$ and $\mathbf{f}^{(t)}_i\!\in\!\mathbb{R}^{d_t}$ which we linearly project to $d$ and $\ell_2$-normalize:
\begin{equation}
\tilde{\mathbf{v}}_i=\mathrm{norm}\!\big(\mathbf{f}^{(v)}_i\mathbf{W}_v\big),\qquad
\tilde{\mathbf{t}}_i=\mathrm{norm}\!\big(\mathbf{f}^{(t)}_i\mathbf{W}_t\big),
\end{equation}
using learnable weights $\mathbf{W}_v\in\mathbb{R}^{d_v\times d}$ and $\mathbf{W}_t\in\mathbb{R}^{d_t\times d}$.

\paragraph{Per-modality $k$NN graphs and frozen fusion.}
We build a cosine-kNN graph among items per modality $m\!\in\!\{v,t\}$. $\hat{\mathbf{A}}_{ii}^{(m)}$ is the adjacency matrix with ones on kNN edges. Using standard GCN normalisation:
\begin{equation}
\mathbf{A}_{ii}^{(m)} \;=\; \mathbf{D}^{-1/2}\hat{\mathbf{A}}_{ii}^{(m)}\mathbf{D}^{-1/2},\qquad
\mathbf{D}=\mathrm{diag}\!\big(\hat{\mathbf{A}}_{ii}^{(m)}\mathbf{1}\big).
\end{equation}
We then fuse the two modality graphs with a \emph{static} weighting $\alpha_v$:
\begin{equation}
\mathbf{A}_{ii}\;=\;\alpha_v\mathbf{A}_{ii}^{(v)} + (1-\alpha_v)\,\mathbf{A}_{ii}^{(t)},\qquad \alpha\in[0,1].
\end{equation}

\paragraph{ID-conditioned gating.}
In contrast to previous works, we introduce a learnable gate that modulates
\emph{how much} text and video features to mix, on a per-item basis.
Importantly, conditioning this gate on the ID space indirectly mitigates the issue of cross-modal misalignment where one modality suggests two items are similar but other modalities disagree. Modality signals that agree with ID signals are amplified, while ones that do not are suppressed.

Let $\mathbf{E}\in\mathbb{R}^{N\times d}$ be the item ID table, and let $\mathbf{e}^i$ denote row $i$.
We compute a \emph{per-item gate} $\beta_i\in[0,1]$ from $\mathbf{e}^i$:
\begin{equation}
\beta_i \;=\; \sigma\!\big(\mathbf{w}_\beta^{\top}\mathbf{e}^i\big),\qquad
\mathbf{w}_\beta\in\mathbb{R}^{d}.
\end{equation}
The gate blends normalized text and image projections to form an ID-anchored content vector:
\begin{equation}
\mathrm{mix}_i \;=\; (1-\beta_i)\,\tilde{\mathbf{t}}_i + \beta_i\,\tilde{\mathbf{v}}_i,
\end{equation}
This makes it possible for certain items to depend more or less heavily on text or visual features. For example, it might be reasonable to expect that the similarity of clothing items is determined more by their visual appearance than their textual description. Meanwhile, the similarity of power tools may be determined more by the description of their function than by their visual appearance.

Finally, we use these gated multimodal embeddings to \emph{seed} the multimodal branch together with $\mathbf{e}^i$ at item $i$ by
\begin{equation}
\mathbf{z}^{(0)}_i \;=\; \mathbf{e}^i \;+\; \alpha_{\mathrm{seed}}\,\mathrm{mix}_i,
\end{equation}
where $\alpha_{\mathrm{seed}}\!\ge\!0$ sets the overall content strength.
This makes multimodal content a residual, ID-anchored cue rather than a replacement for IDs with $\alpha_{\mathrm{seed}}$ constraining its influence. Empirically, small $\alpha_{\mathrm{seed}}$ (0.1 or 0.001 dataset-dependent) works best, which is consistent with treating multimodal content as a calibration aid.

\paragraph{Propagation on the frozen graph.}
We propagate $\mathbf{Z}_{ii}$
over the fused, static adjacency $\mathbf{A}_{ii}$ for $L_{\mathrm{mm}}$ steps with the same neighborhood averaging rule used elsewhere:
\begin{equation}
\mathbf{Z}_{ii}^{(\ell+1)} \;=\; \mathbf{A}_{ii}\,\mathbf{Z}_{ii}^{(\ell)},\qquad \ell=0,\ldots,L_{\mathrm{mm}}-1.
\end{equation}

\paragraph{Output and fusion.}
We denote the multimodal branch output by $\mathbf{e}^{i}_{\mathrm{mm}}=\mathbf{z}^{(L_{\mathrm{mm}})}_i$.
Later, the model fuses $\mathbf{e}^{i}_{\mathrm{mm}}$ with $\mathbf{e}^{\,i}_{\mathrm{ui}}$ using a fixed scalar weight $\alpha_{\mathrm{mm}}$.
\begin{equation}
\tilde{\mathbf{e}}^{i}
=\,\mathbf{e}^{i}_{\mathrm{ui}}
+\alpha_{\mathrm{mm}}\,\mathbf{e}^{i}_{\mathrm{mm}},\quad \lambda_{\mathrm{mm}}\in[0,1],
\end{equation}

\subsection{Training and Optimisation}

\paragraph{Fused scoring.}

We score candidate items for user $u$ by the inner product of $\tilde{\mathbf{e}}^{\,i}$ the fused item vector
and $\mathbf{e}^{\,i}_{\mathrm{ui}}$ the UI-branch item vector
\begin{equation}
\hat{y}_{u,i} \;=\; \langle {\mathbf{e}}^{u}_{\mathrm{ui}},\,\tilde{\mathbf{e}}^{i}\rangle .
\label{eq:score}
\end{equation}

\paragraph{Pairwise ranking objective (BPR)}
Training is driven by a pairwise ranking loss over triples
$\mathcal{T}=\{(u,i^+,i^-)\}$, where $i^+$ is observed for $u$ and $i^-$ is an unobserved negative
sample (drawn uniformly, or randomised within the batch):
\begin{equation}
\mathcal{L}_{\mathrm{BPR}}
\;=\;
-\frac{1}{|\mathcal{T}|}\sum_{(u,i^+,i^-)\in\mathcal{T}}
\log \sigma\!\Big(\hat{y}_{u,i^+} - \hat{y}_{u,i^-}\Big),
\label{eq:bpr}
\end{equation}
with $\sigma(\cdot)$ the logistic sigmoid. Eq.~\eqref{eq:bpr} directly encourages
$\hat{y}_{u,i^+}\!>\!\hat{y}_{u,i^-}$ under implicit feedback.

\paragraph{Cross-view alignment (user-item $\leftrightarrow$ sequence-item).}
We utilise an InfoNCE-style contrastive loss to align matching entities across views, between each user vector and their
sequence node (from the SI branch) and between item representations from the two views. ${\mathbf{s}}^{u}_{si}$ denotes the final sequence
representation for user $u$, and define a cosine similarity
$\mathrm{sim}(\mathbf{a},\mathbf{b})
=\big\langle \frac{\mathbf{a}}{\|\mathbf{a}\|},\,\frac{\mathbf{b}}{\|\mathbf{b}\|}\big\rangle$.
With temperature $\tau\!>\!0$ and a negative set $\mathcal{N}(u)$ of other users’
sequences, the loss is defined as
\begin{equation}
\mathcal{L}_{\mathrm{US}}
=
-\frac{1}{|\mathcal{U}|}\sum_{u\in\mathcal{U}}
\log
\frac{\exp\!\big(\mathrm{sim}(\mathbf{e}^{u}_{\mathrm{ui}},\,\mathbf{s}^{u}_{\mathrm{si}})/\tau\big)}
{\sum\limits_{u'\in\mathcal{U}}
\exp\!\big(\mathrm{sim}(\mathbf{e}^{u}_{\mathrm{ui}},\,\mathbf{s}^{u'}_{\mathrm{si}})/\tau\big)} .
\label{eq:us-compact}
\end{equation}

For the positive items in the batch, with $\tilde{\mathbf{e}}^{\,i}$ the fused item vector
and $\mathbf{e}^{\,i}_{\mathrm{si}}$ the SI-branch item vector,
\begin{equation}
\mathcal{L}_{\mathrm{IS}}
=
-\frac{1}{|\mathcal{B}|}\sum_{i\in\mathcal{B}}
\log
\frac{\exp\!\big(\mathrm{sim}(\tilde{\mathbf{e}}^{\,i},\,\mathbf{e}^{\,i}_{\mathrm{si}})/\tau\big)}
{\sum\limits_{i'\in\mathcal{B}}
\exp\!\big(\mathrm{sim}(\tilde{\mathbf{e}}^{\,i},\,\mathbf{e}^{\,i'}_{\mathrm{si}})/\tau\big)}.
\label{eq:item-seq}
\end{equation}

\paragraph{Multimodal alignment loss.}
We define a \emph{modality-aware} pairwise loss that aligns the UI-branch user vector ${\mathbf{e}}^{u}_{\mathrm{ui}}$ with the corresponding content spaces:
\begin{equation}
\label{eq:lmm}
\mathcal{L}_{\mathrm{MM}}
=
-\frac{1}{|\mathcal{T}|}
\sum_{(u,i^+,i^-)\in\mathcal{T}}
\;\sum_{m\in\{t,v\}}
\log \sigma\!\left(
\big\langle \mathbf{e}^{u}_{\mathrm{ui}},\,\tilde{\mathbf{m}}_{i^+}-\tilde{\mathbf{m}}_{i^-}\big\rangle
\right),
\end{equation}

\paragraph{Regularisation and total objective.}
The complete loss combines supervised ranking, optional cross-view alignment, and weight
regularisation:
\begin{equation}
\mathcal{L}
\;=\;
\mathcal{L}_{\mathrm{BPR}}
\;+\;
\lambda_{u}\,\mathcal{L}_{\mathrm{US}}
\;+\;
\lambda_{i}\,\mathcal{L}_{\mathrm{IS}}
\;+\;
\lambda_{\mathrm{sv}}\,\mathcal{L}_{\mathrm{SV}}
\;+\;
\lambda_{\mathrm{mm}}\,\mathcal{L}_{\mathrm{MM}},
\label{eq:total}
\end{equation}
where $\lambda_{u}, \lambda_{i}, \lambda_{\mathrm{sv}}$ and $\lambda_{\mathrm{mm}}$ are non-negative scalar weights.

\section{Experiments}

\begin{table*}[t]
\caption{Overall performance comparison of the different recommendation models. The best result on each dataset for each metric is boldened, and the second best is underlined.}
\centering
\small
\begin{tabular}{l cccc cccc cccc}
\toprule
\textbf{Datasets} & \multicolumn{4}{c}{\textbf{Baby}} & \multicolumn{4}{c}{\textbf{Sports}} & \multicolumn{4}{c}{\textbf{Elec}} \\
\cmidrule(lr){2-5} \cmidrule(lr){6-9} \cmidrule(lr){10-13}
\textbf{Metrics} & \textbf{R@10} & \textbf{R@20} & \textbf{N@10} & \textbf{N@20} & \textbf{R@10} & \textbf{R@20} & \textbf{N@10} & \textbf{N@20} & \textbf{R@10} & \textbf{R@20} & \textbf{N@10} & \textbf{N@20} \\
\midrule
\textbf{LightGCN} & 0.0336 & 0.0549 & 0.0178 & 0.0231 & 0.0402 & 0.0631 & 0.0213 & 0.0270 & 0.0288 & 0.0427 & 0.0154 & 0.0189 \\
\textbf{SGL} & 0.0361 & 0.0586 & 0.0188 & 0.0244 & 0.0425 & 0.0688 & 0.0227 & 0.0293 & 0.0292 & 0.0439 & 0.0158 & 0.0195 \\ \midrule
\textbf{SASRec} & 0.0298 & 0.0478 & 0.0150 & 0.0195 & 0.0274 & 0.0421 & 0.014 & 0.0177 & 0.0241 & 0.0368 & 0.0128 & 0.0160 \\
\textbf{BERT4Rec} & 0.0207 & 0.0355 & 0.0102 & 0.0139 & 0.0176 & 0.0273 & 0.0094 & 0.0118 & 0.023 & 0.0358 & 0.0117 & 0.0149 \\
\textbf{FEARec} & 0.0285 & 0.048 & 0.0146 & 0.0195 & 0.0281 & 0.0456 & 0.0143 & 0.0187 & 0.0255 & 0.039 & 0.0136 & 0.0170 \\
\textbf{SRGNN} & 0.0260 & 0.0418 & 0.0132 & 0.0172 & 0.0196 & 0.0314 & 0.0102 & 0.0132 & 0.0242 & 0.0367 & 0.0128 & 0.0159 \\ \midrule
\textbf{MMGCN} & 0.0253 & 0.0436 & 0.0124 & 0.0169 & 0.0275 & 0.0447 & 0.014 & 0.0183 & 0.0173 & 0.0285 & 0.0086 & 0.0114 \\
\textbf{GRCN} & 0.0359 & 0.0574 & 0.0191 & 0.0245 & 0.0371 & 0.0606 & 0.0189 & 0.0249 & 0.0232 & 0.0367 & 0.0121 & 0.0155 \\
\textbf{VBPR} & 0.0313 & 0.0517 & 0.0163 & 0.0214 & 0.0406 & 0.0638 & 0.0204 & 0.0263 & 0.0228 & 0.0348 & 0.0113 & 0.0143 \\
\textbf{BM3} & 0.0371 & 0.0612 & 0.0189 & 0.0250 & 0.0435 & 0.0698 & 0.0224 & 0.0290 & 0.0302 & 0.0450 & 0.0164 & 0.0201 \\
\textbf{MGCN} & 0.0420 & 0.0666 & 0.0222 & 0.0282 & 0.0476 & 0.0750 & 0.0251 & 0.0320 & 0.0330 & 0.0495 & 0.0179 & 0.0221 \\
\textbf{FREEDOM} & 0.0411 & 0.0655 & 0.0210 & 0.0272 & 0.0479 & \underline{0.0752} & 0.0246 & 0.0315 & 0.0315 & 0.0482 & 0.0169 & 0.0211 \\
\textbf{LGMRec} & 0.0412 & 0.0649 & 0.0212 & 0.0271 & 0.0455 & 0.0715 & 0.0238 & 0.0303 & 0.0328 & 0.0496 & 0.0176 & 0.0219 \\
\textbf{SMORE} & \underline{0.0439} & \underline{0.0684} & \underline{0.0229} & \underline{0.0291} & \underline{0.0480} & 0.0746 & \underline{0.0254} & \underline{0.0321} & \underline{0.0333} & \underline{0.0499} & \underline{0.0181} & \underline{0.0222} \\ \midrule
\textbf{MuSICRec*} & \textbf{0.0455} & \textbf{0.0718} & \textbf{0.0235} & \textbf{0.0300} & \textbf{0.0508} & \textbf{0.0805} & \textbf{0.0267} & \textbf{0.0341} & \textbf{0.0350} & \textbf{0.0519} & \textbf{0.0189} & \textbf{0.0232} \\
\textbf{Improvement} & 3.64\% & 4.97\% & 2.62\% & 3.09\% & 5.83\% & 7.05\% & 5.12\% & 6.23\% & 5.11\% & 4.01\% & 4.42\% & 4.50\% \\
\bottomrule
\end{tabular}
\label{tab:joint-results}
\end{table*}

We perform multiple experiments to evaluate our proposed model's performance and benchmark it against a broad range of existing state-of-the-art recommendation systems, including generic, multimodal, and sequential methods in our analysis.

A key challenge in this breadth arises from the different evaluation protocols used in sequential and multimodal recommendation research, for example, variations in train-validation-test data split and metric definition. Consequently, although we are comparing on standard datasets, we were forced to develop a more unified evaluation protocol to enable fair and meaningful comparisons across the broad range of distinct approaches.

In addition to benchmarking against prior work, we conduct an ablation study to determine the contribution of individual system components. We also analyse the sensitivity of MuSICRec with respect to \(\lambda_{u}\) and \(\lambda_{i}\), which control the alignment of the user-item and sequence-item views.

\subsection{Experimental Setup}
\label{sec:experimental_setup}
Multimodal recommender systems are atemporal and typically rely on random train/test splits. This does not mesh well with sequential recommendation, where the prediction target must lie in the future relative to the training data. We therefore define a unified evaluation protocol where we use all interactions except the last two for training, keeping the penultimate and final interactions for validation and testing, respectively. We deviate from the typical leave-one-out split usually used to evaluate sequential recommendation by not feeding the validation item back into the test prefix. This means that we provide the same exact prefix for validation and testing, which only includes the training interactions. This is done because in multimodal and generic recommendation systems, the validation interaction is kept out and not used to build or update the user-item graph before testing. In other words, there is no message passing or representation updates that use the penultimate validation item. Similarly, in MuSICRec, we neither update the user-item adjacency matrix nor the sequence-item adjacency matrix or initial sequence nodes (formed by attention pooling over item ID embeddings forming user history) with the validation item. This avoids leakage by ensuring information from held-out interactions can not slip into test item representation through graph updates and ensures an equitable comparison between all baselines under the same strict leave-two-out split. Our choice is consistent with the practice of building multimodal recommenders on fixed interaction graphs and recent cautions about leakage in offline recommender evaluation \cite{ji2023critical}. All baselines were rerun with this leave-two-out split, and thus, numbers may differ slightly from those reported in their respective publications.

\subsection{Datasets}
\begin{table}[htbp]
\centering
\caption{Datasets Statistics}
\vskip 0.1in
\label{tab:amazon_5core_stats}
\begin{tabular}{lrrrr}
\toprule
Dataset & Users & Items & Reviews & Sparsity (\%) \\
\midrule
\textit{Baby}        & 19{,}445   & 7{,}333    & 160{,}792   & 99.89 \\
\textit{Sports}      & 35{,}598   & 18{,}357   & 296{,}337   & 99.95 \\
\textit{Elec}        & 192{,}403  & 63{,}001   & 1{,}689{,}188 & 99.99 \\
\bottomrule
\end{tabular}
\end{table}

The publicly available Amazon datasets are widely used to evaluate sequential and multimodal recommender systems. These contain product reviews extracted by Amazon.com and are among the few datasets that provide textual and visual data that could be utilised for our task. We focus on the Baby, ``Sports and Outdoors'' and Electronics datasets in our evaluation, and refer to them as \textbf{Baby}, \textbf{Sports} and \textbf{Elec}. The text and image URLs are provided with the metadata for Amazon datasets. We utilise the published 384-dimensional text features and 4096-dimensional visual features. 5-core filtering of users and items is carried out on all three datasets and user ratings are treated as positive interactions.

Recall (R@k) and Normalised Discounted Cumulative Gain (N@k) are the two metrics we report in our evaluation. Under our leave-two-out protocol, the validation and test sets comprise one item per user. R@k indicates whether the ground-truth item appears in the top-k list, and N@k is rank-sensitive, applying a logarithmic discount to the position of the correct item, thus rewarding higher placement.

\subsection{Baselines}
We compare against state-of-the-art multimodal and sequential baselines. These baselines fall under these categories: 
\begin{itemize}
    \item ID-only collaborative methods: LightGCN\cite{he2020lightgcn} and SGL\cite{Qiu_2022}
    \item Multimodal models: VBPR\cite{he2016vbpr}, MMGCN\cite{wei2019mmgcn}, MGCN\cite{Yu_2023}, GRCN \cite{yu2021graph}, FREEDOM\cite{zhou2023tale}, LGMRec\cite{guo2024lgmreclocalglobalgraph}, BM3\cite{zhou2023bootstrap} and SMORE\cite{SMORE}
    \item sequential baselines: SASRec\cite{kang2018self}, BERT4Rec\cite{sun2019}, FEARec\cite{du2023frequencyenhancedhybridattention}, and SR-GNN\cite{Wu_2019_srgnn}.
\end{itemize}

\subsection{Implementation Details}
MuSICRec and all baselines were implemented in PyTorch, trained with an Adam optimiser, Xavier initialisation, and fixed user item and sequence dimensionality to 64. For MuSICRec, we set the number of GCN layers in the item-item graph at $L_{ii} = 1$, the user-item graph at $L_{ui} = 3$, and the sequence-item graph at $L_{si} = 2$. We set $\alpha_v=0.1$, vary $\alpha_{seed}$ in $\{0.001, 0.01, 0.1\}$ and cap the maximum sequence length at $L_{\max}=50$. We train up to 1000 epochs with early stopping (patience) set to 20, using validation R@20 as the training stopping indicator. MuSICRec, LightGCN, and all multimodal baselines are implemented within the unified MMRec framework. All sequential baselines and SGL are trained with the RecBole library \cite{zhao2021recboleunifiedcomprehensiveefficient}. MMRec is based on and integrates with RecBole. Both evaluation pipelines were made to align with each other, setting training and valid/test batch size to 512 and 4096, respectively, and using the leave-two-out split described in Section \ref{sec:experimental_setup}. Hyperparameter sweeps for baselines follow their paper's recommendation, and the best test results, based on valid R@20, are reported for all models. All baselines and MuSICRec use BPR loss as the primary objective for optimisation.

\subsection{Performance Comparison}
Table \ref{tab:joint-results} reports R@10 and @20 and N@10 and @20 on \textbf{Baby}, \textbf{Sports} and \textbf{Elec}. MuSICRec achieves the best results across all datasets and metrics.
On \textbf{Baby}, MuSICRec surpasses the best baseline by an average across all metrics of 3.58\%. Similarly, average improvements of 6.05\% and 4.51\% were obtained on \textbf{Sports} and \textbf{Elec}, respectively. The best baselines are consistently multimodal, with larger gains obtained over sequential models and LightGCN and SGL. Averaged over all datasets and metrics, MuSICRec improves by 4.72\% relative to the next best result. 

These results show that MuSICRec, regardless of dataset size and sparsity, consistently outperforms state-of-the-art multimodal and sequential recommender systems. These gains indicate the utility of building and propagating signals from the sequence-item graph and passing ID-conditioned mix of multimodal information into the frozen item-item graph, leading to consistent gains. These remain beneficial even at a larger scale as seen on \textbf{Elec}.

\subsection{Ablation Study}
\label{ablation}
To isolate the contribution of MuSICRec's individual components, we conduct an ablation study evaluating the following variations of the model:
\noindent\textbf{(i) w/o UI graph} removes user-item graph contribution\\
\textbf{(ii) w/o SI graph} removes sequence-item graph\\
\textbf{(iii) w/o MM} removes multimodal item-item graph\\
\textbf{(iv) w/o MM-ID seed} disables ID-guided multimodal seeding\\
\textbf{(v) w/ Sinusoidal Pooling} replaces learned sequence attention pooling with a parameter-free pooling that adds fixed sinusoidal positional encodings to item embeddings before aggregation.\\
As shown in Figures \ref{fig:sports_ablation} and Figures \ref{fig:baby_ablation}, and \ref{fig:elec_ablation} in Appendix \ref{appendix},
dropping UI graph propagation and aggregation yields the largest degradation on all datasets. Multimodal and sequential signals rely on and complement collaborative behaviour relationships; thus, the user-item graph can not be fully substituted. 
Removing SI consistently hurts performance, showing that pooling sequence representations into the graph adds orthogonal information that complements the UI graph.
\textbf{w/o MM} reduces accuracy as would be expected from removing item multimodal contribution entirely.
\textbf{w/o MM-ID seed} performs below the full model but above \textbf{w/o MM}, indicating that seeding and gating the multimodal information with item IDs prior to item-item graph propagation improves alignment and denoising. Sinusoidal pooling underperforms learned attention pooling, demonstrating that data-dependent, learnable fusion is needed to decide how much each interaction (recent vs. early, bursty vs. steady) should influence the sequence representation prior to graph propagation.

\begin{figure}[!htbp]
\centering
\includegraphics[width=0.8\columnwidth]{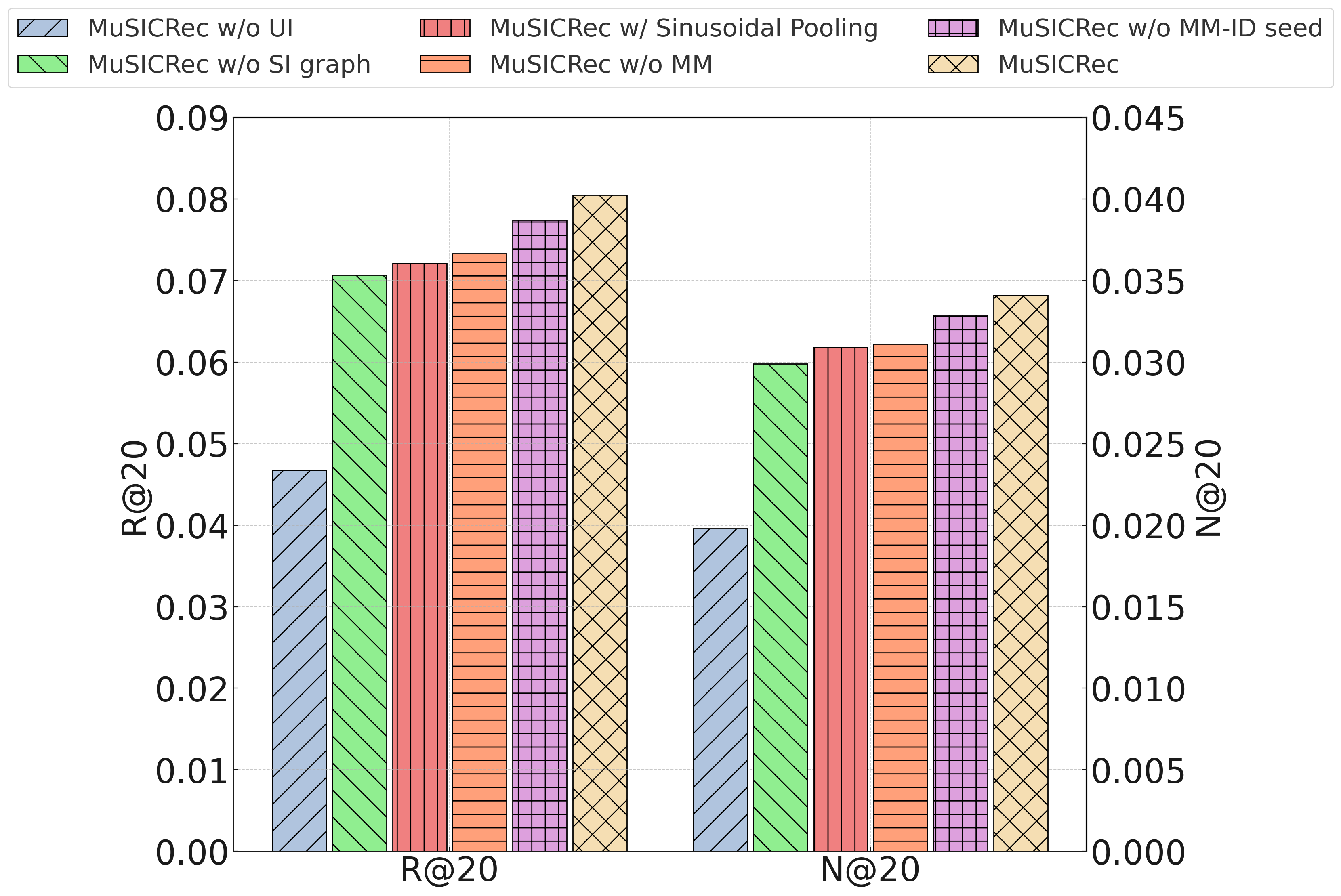}
\caption{\emph{Sports} Ablation}
\label{fig:sports_ablation}
\end{figure}

\subsection{Sensitivity Study}

\begin{figure}[!htbp]
\centering
\includegraphics[width=0.8\columnwidth]{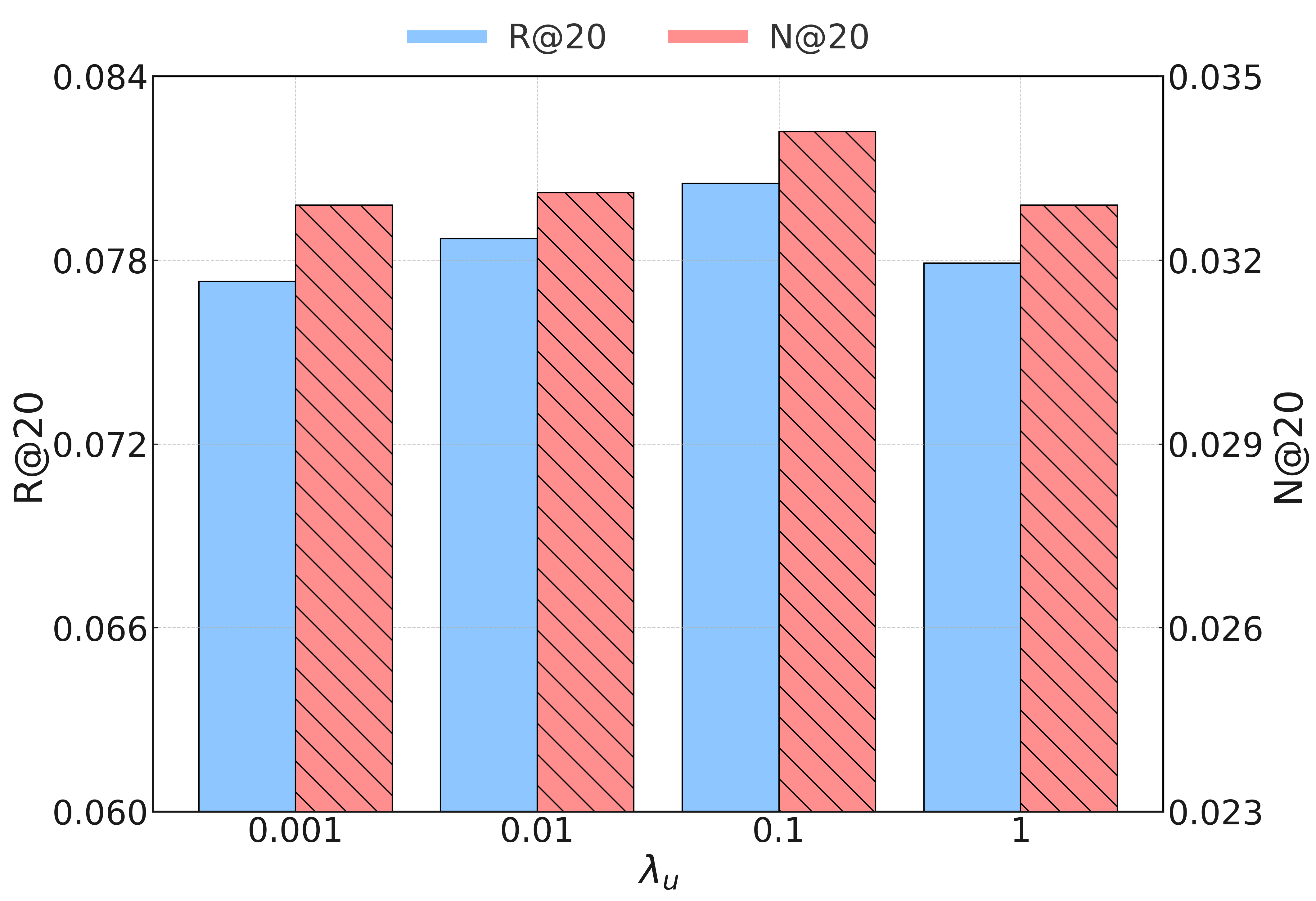}
\caption{$\lambda_u$ \emph{Sports} Sensitivity}
\label{fig:sports_sensitivity_u}
\end{figure}

\begin{figure}[!htbp]
\centering
\includegraphics[width=0.8\columnwidth]{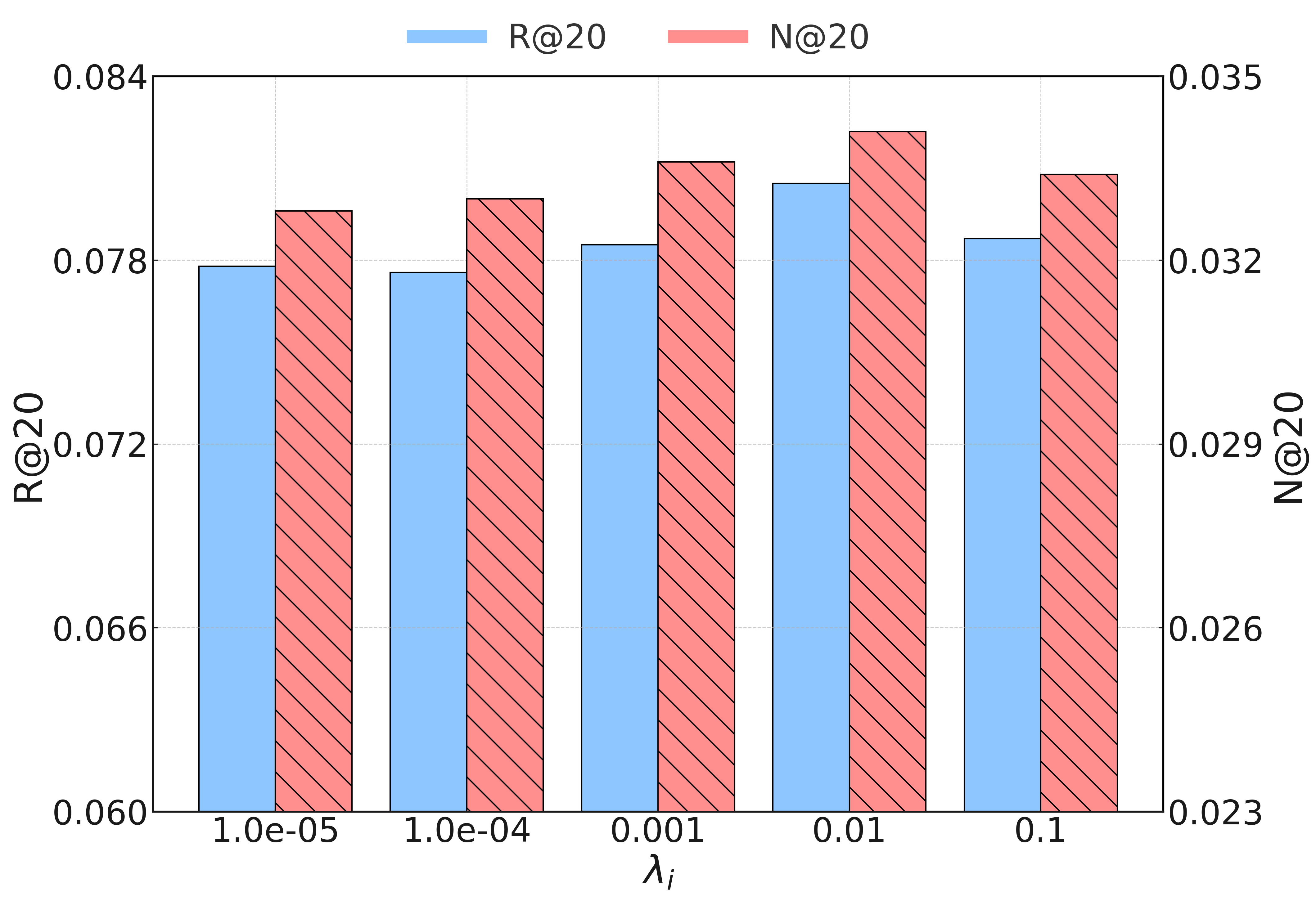}
\caption{$\lambda_i$ \emph{Sports} Sensitivity}
\label{fig:sports_sensitivity_i}
\end{figure}

We analyse $\lambda_{u}$ and $\lambda_{i}$, which control the contribution of the contrastive losses to the overall objective, balancing the input of the sequence-item view, affecting recommendation performance. Figures \ref{fig:sports_sensitivity_u} and \ref{fig:sports_sensitivity_i} and Figures \ref{fig:baby_sensitivit_u}, \ref{fig:baby_sensitivity_i},  \ref{fig:elec_sensitivity_u} and \ref{fig:elec_sensitivity_i} in Appendix \ref{appendix} show the sweeps conducted and the results obtained. MuSICRec performs best when setting $\lambda_{u}$ to 0.1 on \textbf{Sports} and \textbf{Elec} and to 0.001 on \textbf{Baby}. $\lambda_{i}$ is best when set to 0.01 for all datasets.  We also note that parameter sensitivity is low, and performance is generally stable with only mild accuracy drift when setting $\lambda_{u}$ and $\lambda_{i}$ too high or too low.

\subsection{User History Depth Study}
\begin{figure}[!htbp]
\centering
\includegraphics[width=0.8\columnwidth]{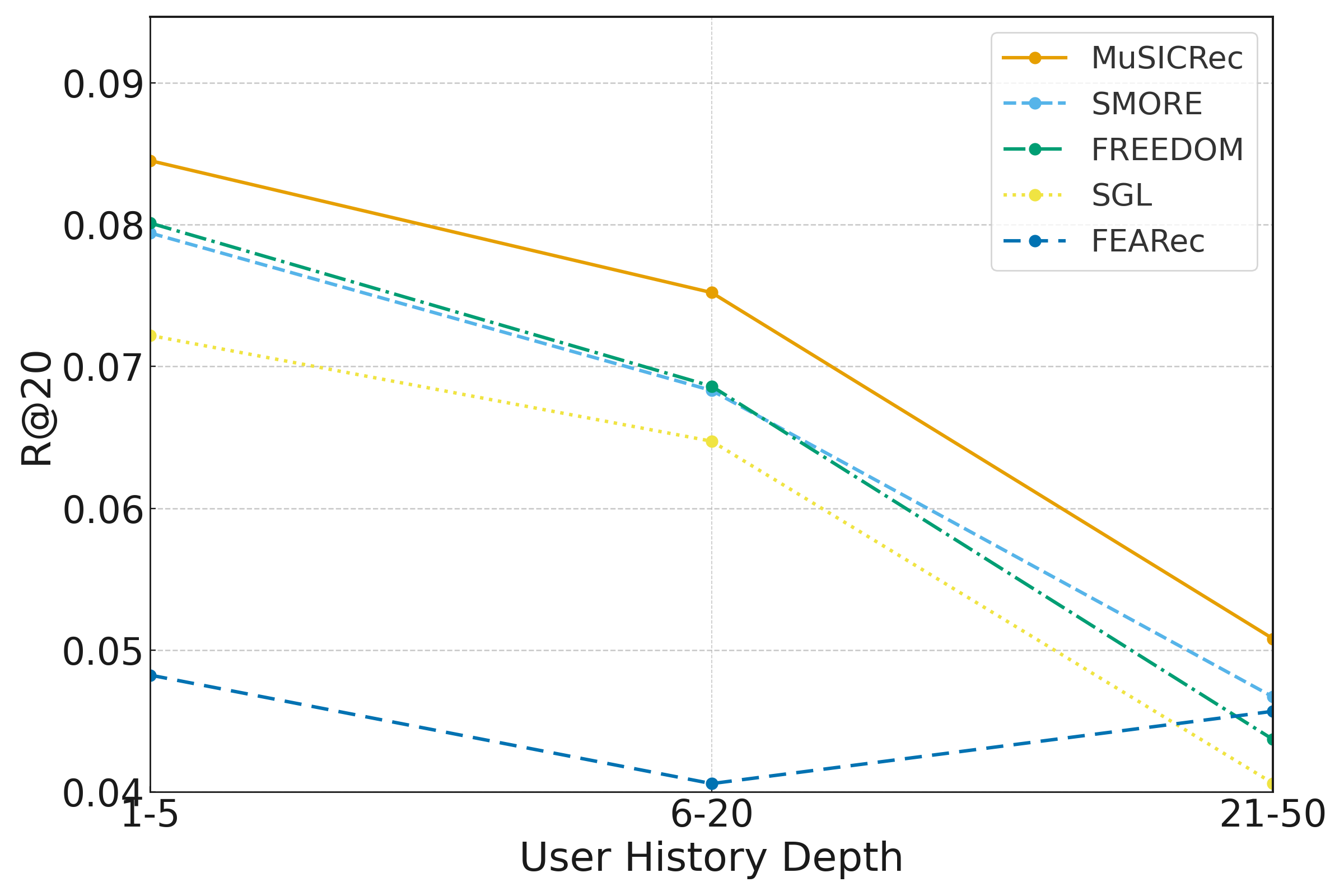}
\caption{\emph{Sports} User History Depth Analysis}
\label{fig:sports_bucket_r20}
\end{figure}
We filter users by history length, with near-cold users having 1-5 interactions, warm users with 6-20 interactions, and long-term users with 21-50 interactions. This bucket-wise evaluation reveals behaviour and cold/near-cold conditions while checking that gains still persist for users with denser histories. As can be seen in Figure \ref{fig:sports_bucket_r20}, MuSICRec delivers the largest gains for the 1-5 bucket, while improving accuracy to a lesser extent for 6-20 and 21-50 users. This shows that the extensive collaborative filtering and information sharing in MuSICRec helps to tackle common cold start and data sparsity issues. 
Similarly, cross and intra-modal denoised multimodal representations also allow MuSICRec to leverage relevant side information when histories are short. 
However, it is pleasing to see that as histories lengthen, performance remains strong and the technique does not merely overfit to short histories, to the detriment of other users. 

\subsection{Computation Cost}
From a complexity standpoint, \textbf{MuSICRec} with $L_{ii} = 1$ $L_{ui} = 3$ $L_{si} = 2$ executes six sparse propagations per epoch and one global sequence-cache refresh. Each sparse propagation costs \(\Theta\!\big(\mathrm{nnz}(\mathbf{A})\cdot d\big)\) for adjacency \(\mathbf{A}\) (SpMM on \([N\times N]\times[N\times d]\)), giving a leading term \(\Theta\!\big((3|L_{\mathrm{ui}}|+2|L_{\mathrm{si}}+1|L_{\mathrm{mm}}|)\cdot d\big)\). The sequence node refresh applies a single-head additive attention over at most \(L_{\max}=50\) tokens per sequence in every batch, adding \(\Theta\!\big(Batch\_Size\cdot L\cdot d\big)\). No transfroms, biases or nonlinearities are applied between layers.

We benchmark computation cost using train epoch time and peak allocated GPU memory over full train$\rightarrow$val$\rightarrow$test for each model.
Table \ref{tab:compute_sports} shows the computation cost results on Sports dataset. Compared to the strongest baseline from Table \ref{tab:joint-results} (SMORE), MuSICRec trains at faster speed per epoch while using lower peak VRAM, despite achieving better metrics. We train at faster epoch time and have lower GPU utilisation compared to most baseline models compared against.


\begin{table}[!htbp]
\caption{Computation cost on \textbf{Sports} with train batch size of 512. This shows pure training epoch time and peak GPU memory allocated over train$\rightarrow$val$\rightarrow$test.}
\centering
\small
\begin{tabular}{l cc}
\toprule
\textbf{Model} & \textbf{Train Epoch Time (s)} & \textbf{Peak GPU Memory (MiB)} \\
\midrule
LightGCN & 6.90 & 681.32 \\
SGL & 16.31 & 355.21 \\
SASRec & 16.53 & 1531.80 \\
BERT4Rec & 23.01 & 1427.60 \\
FEARec & 40.39 & 1774.61 \\
SRGNN & 59.22 & 1302.96 \\
MMGCN & 60.05 & 2033.10 \\
GRCN & 18.96 & 2820.14 \\
VBPR & 2.60 & 1871.70 \\
BM3 & 8.84 & 2700.88 \\
MGCN & 13.16 & 3620.33 \\
FREEDOM & 8.96 & 3638.33 \\
LGMRec & 19.54 & 2711.77 \\
SMORE & 18.52 & 2712.22 \\
MuSICRec & 13.31 & 2043.97 \\
\bottomrule
\end{tabular}
\label{tab:compute_sports}
\end{table}

\section{Conclusion}
In this paper, we proposed a novel model MuSICRec that leverages sequences of interactions as a class of nodes within a new sequence-item graph. It unifies sequential signals from this graph user-item and multimodal item-item signals in a single contastive learning framework. In our extensive experiments, it delivered consistent gains over recent strong baselines. Ablations verified the contribution of each component, and sensitivity analysis showed robustness to $\lambda_u$ and $\lambda_i$. In future work, we aim to improve sequence representation learning to further better performance.

\bibliographystyle{ACM-Reference-Format}
\bibliography{samples/sample-base}

\clearpage

\twocolumn[
\appendix
\section{Additional Figures}
\label{appendix}\label{appendix:figs}

\centering
\captionsetup{type=figure} 

\setcounter{subfigure}{0}
\subcaptionbox{\emph{Baby} ablation\label{fig:baby_ablation}}
  {\includegraphics[width=0.32\textwidth]{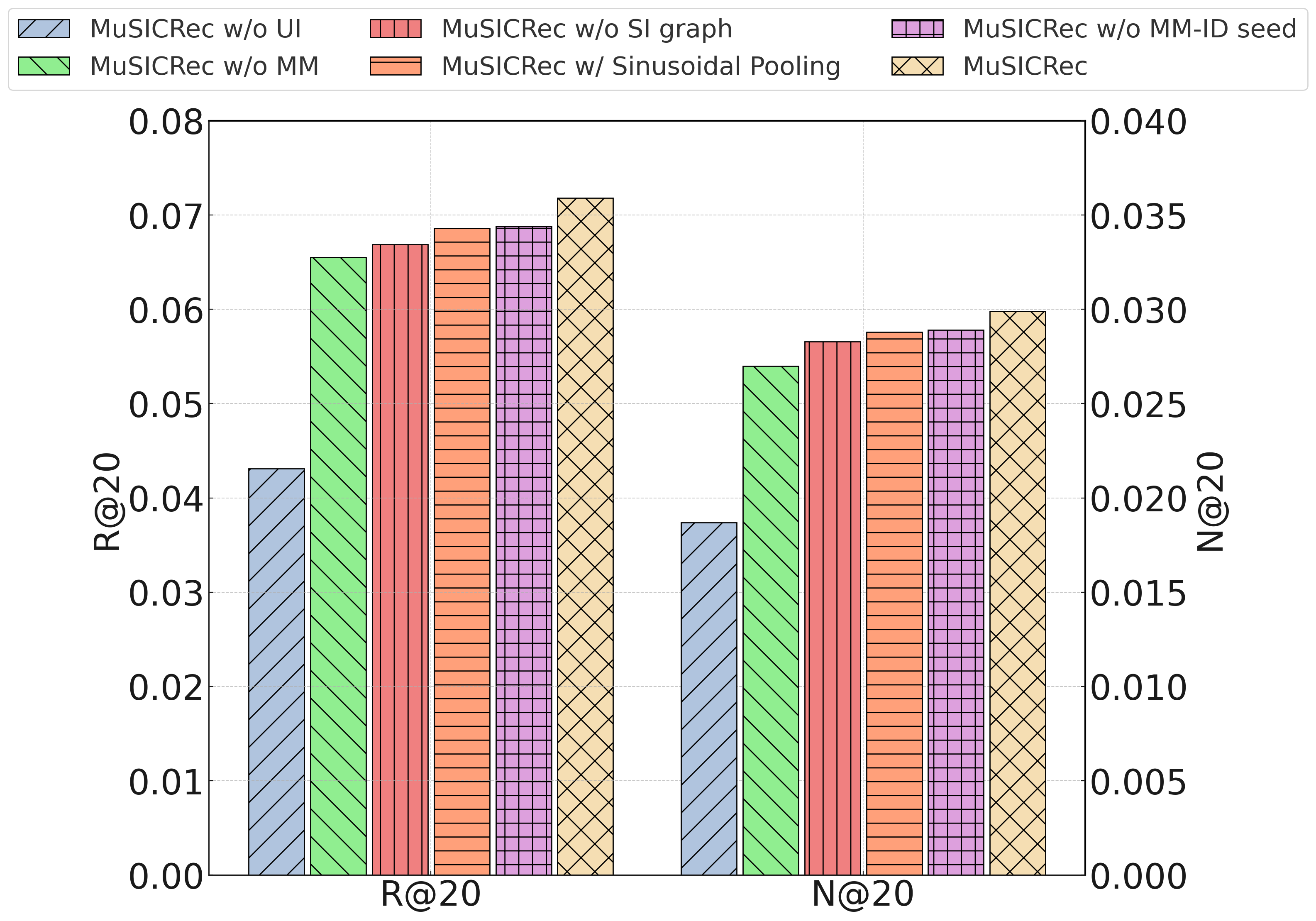}}\hfill
\subcaptionbox{\emph{Elec} ablation\label{fig:elec_ablation}}
  {\includegraphics[width=0.32\textwidth]{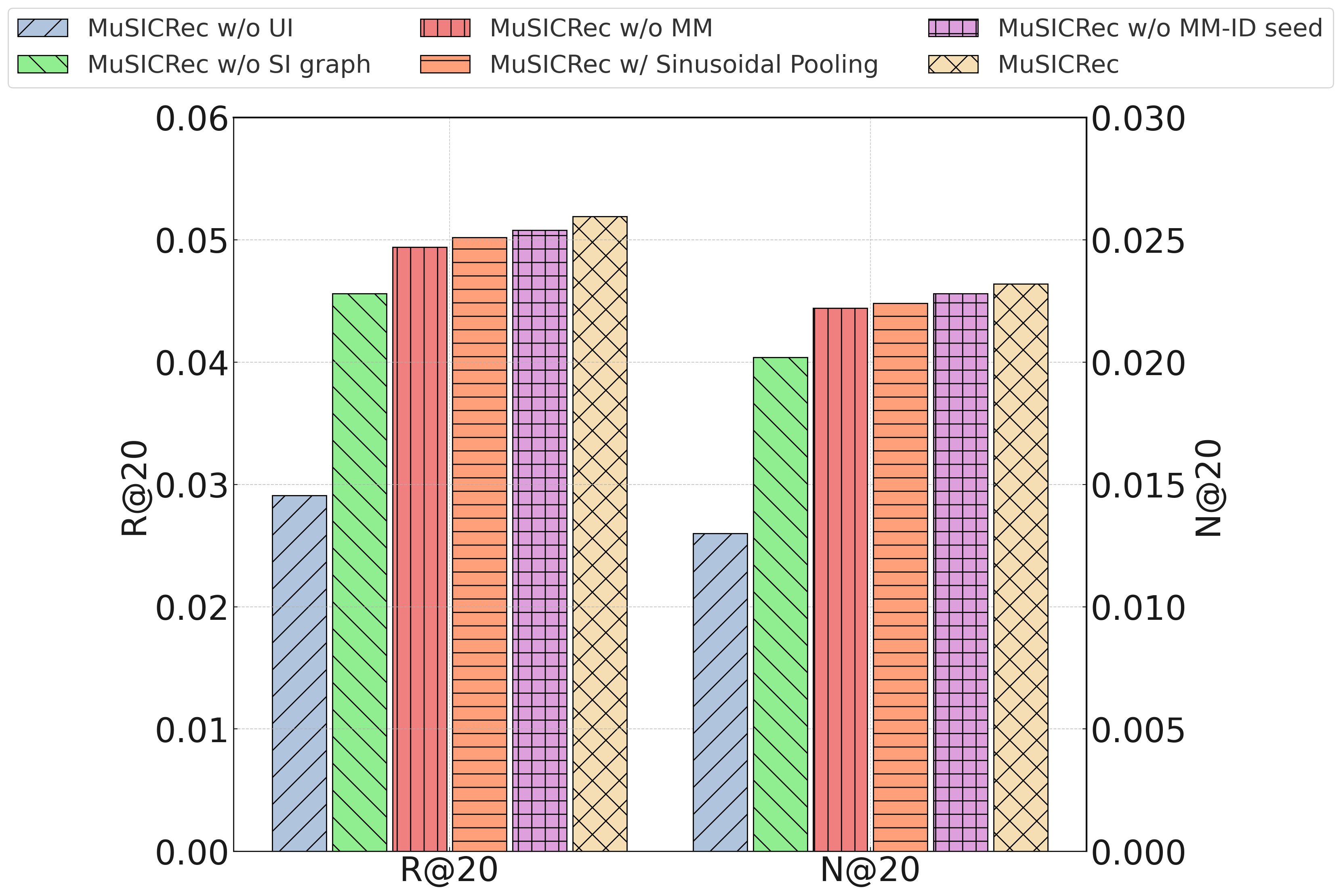}}\hfill
\subcaptionbox{\emph{Baby} $\lambda_u$\label{fig:baby_sensitivit_u}}
  {\includegraphics[width=0.32\textwidth]{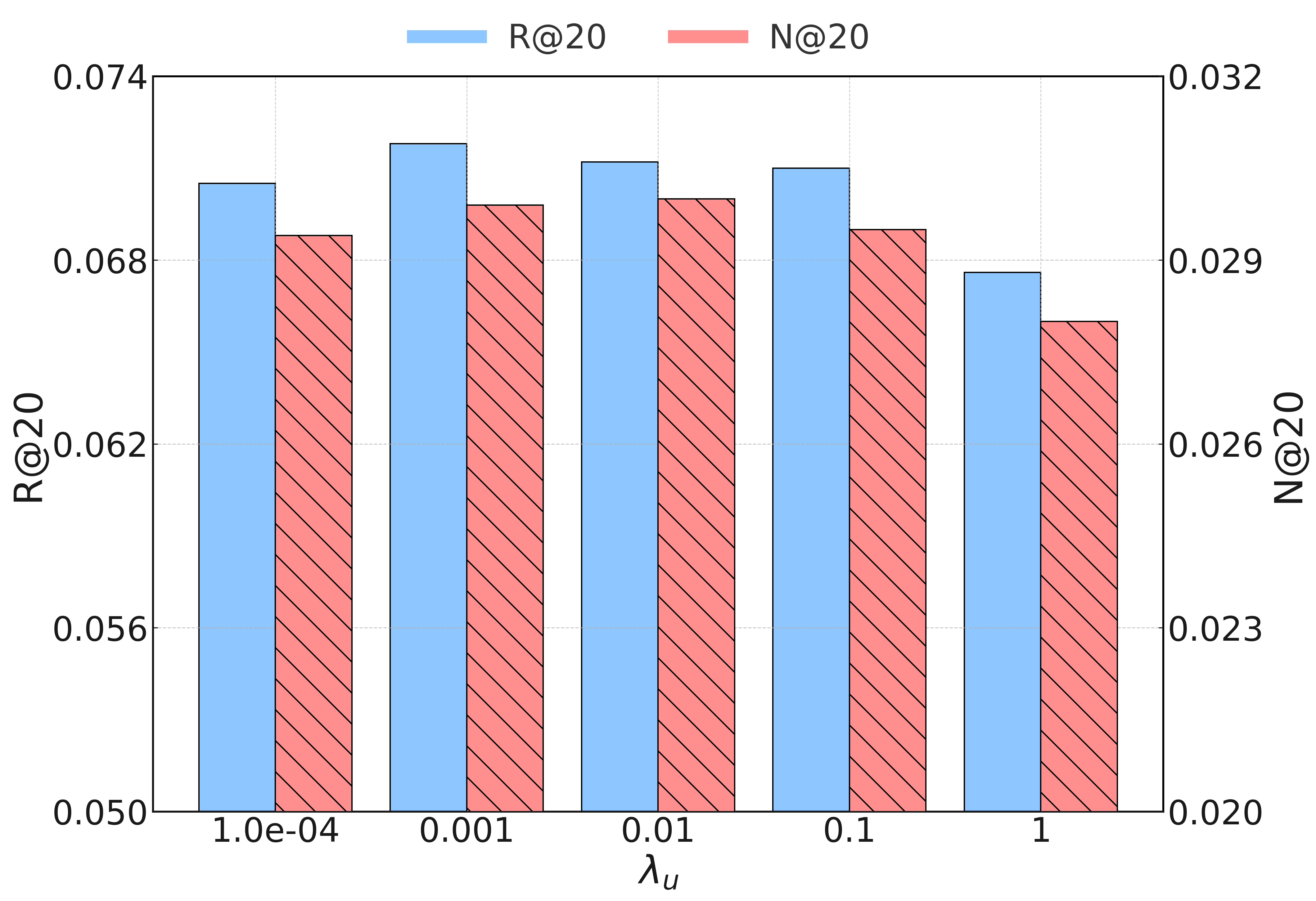}}\\

\subcaptionbox{\emph{Baby} $\lambda_i$\label{fig:baby_sensitivity_i}}
  {\includegraphics[width=0.32\textwidth]{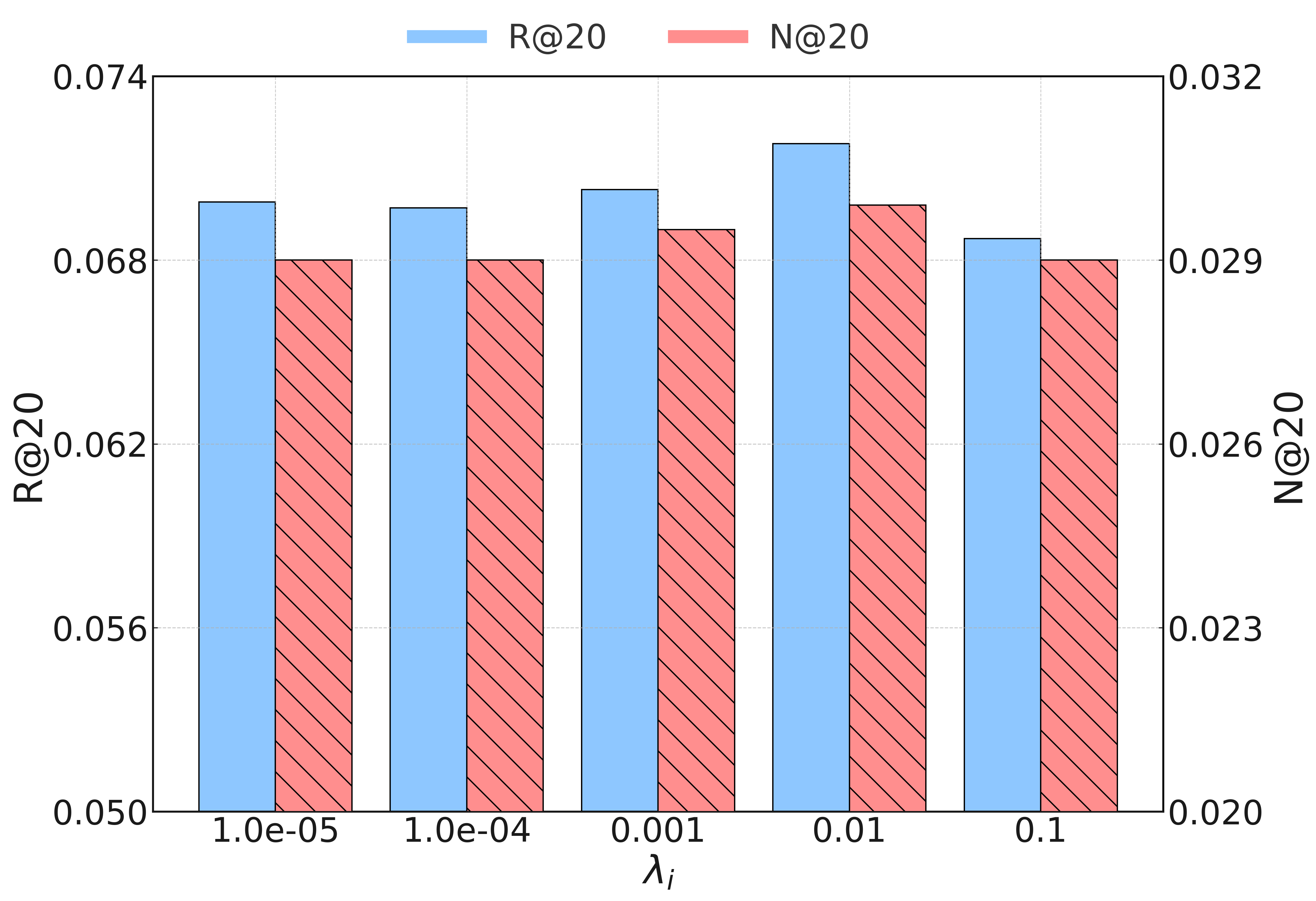}}\hfill
\subcaptionbox{\emph{Elec} $\lambda_u$\label{fig:elec_sensitivity_u}}
  {\includegraphics[width=0.32\textwidth]{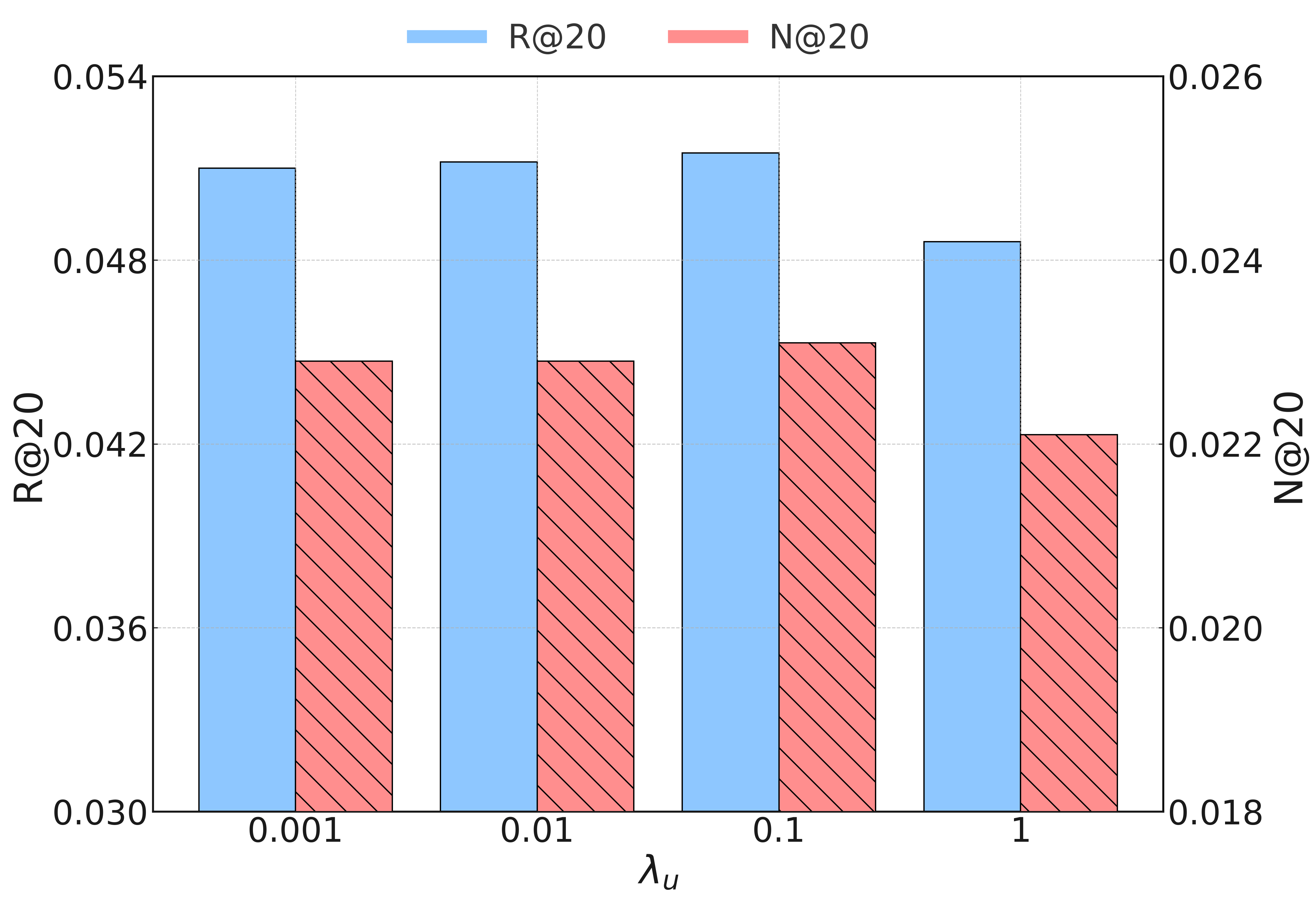}}\hfill
\subcaptionbox{\emph{Elec} $\lambda_i$\label{fig:elec_sensitivity_i}}
  {\includegraphics[width=0.32\textwidth]{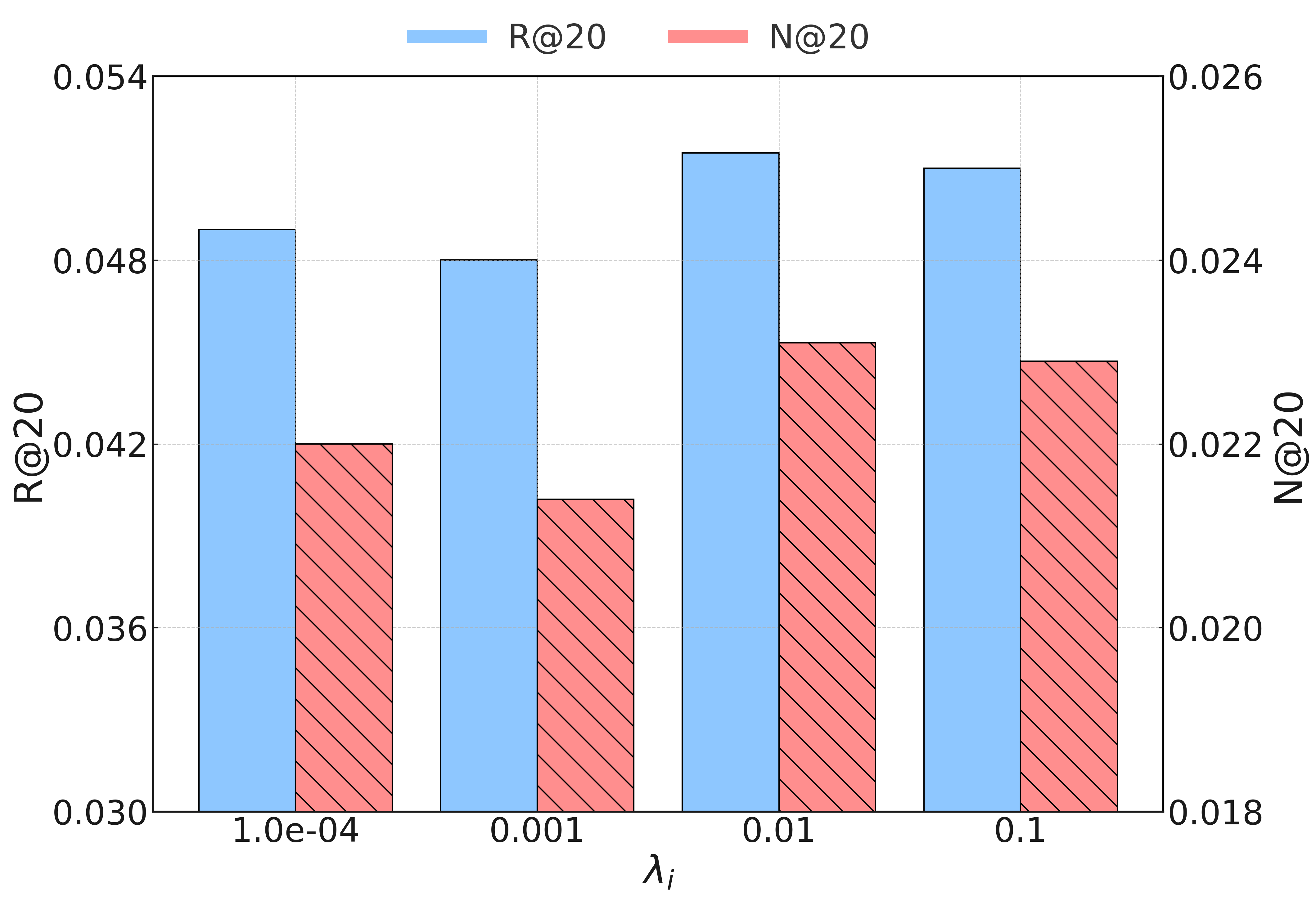}}\\

\captionof{figure}{Ablation and sensitivity analyses (additional results).}
\label{fig:app_ablation_sensitivity}

\vspace{2mm}

\setcounter{subfigure}{0}
\subcaptionbox{\emph{Sports} N@20}{\includegraphics[width=0.32\textwidth]{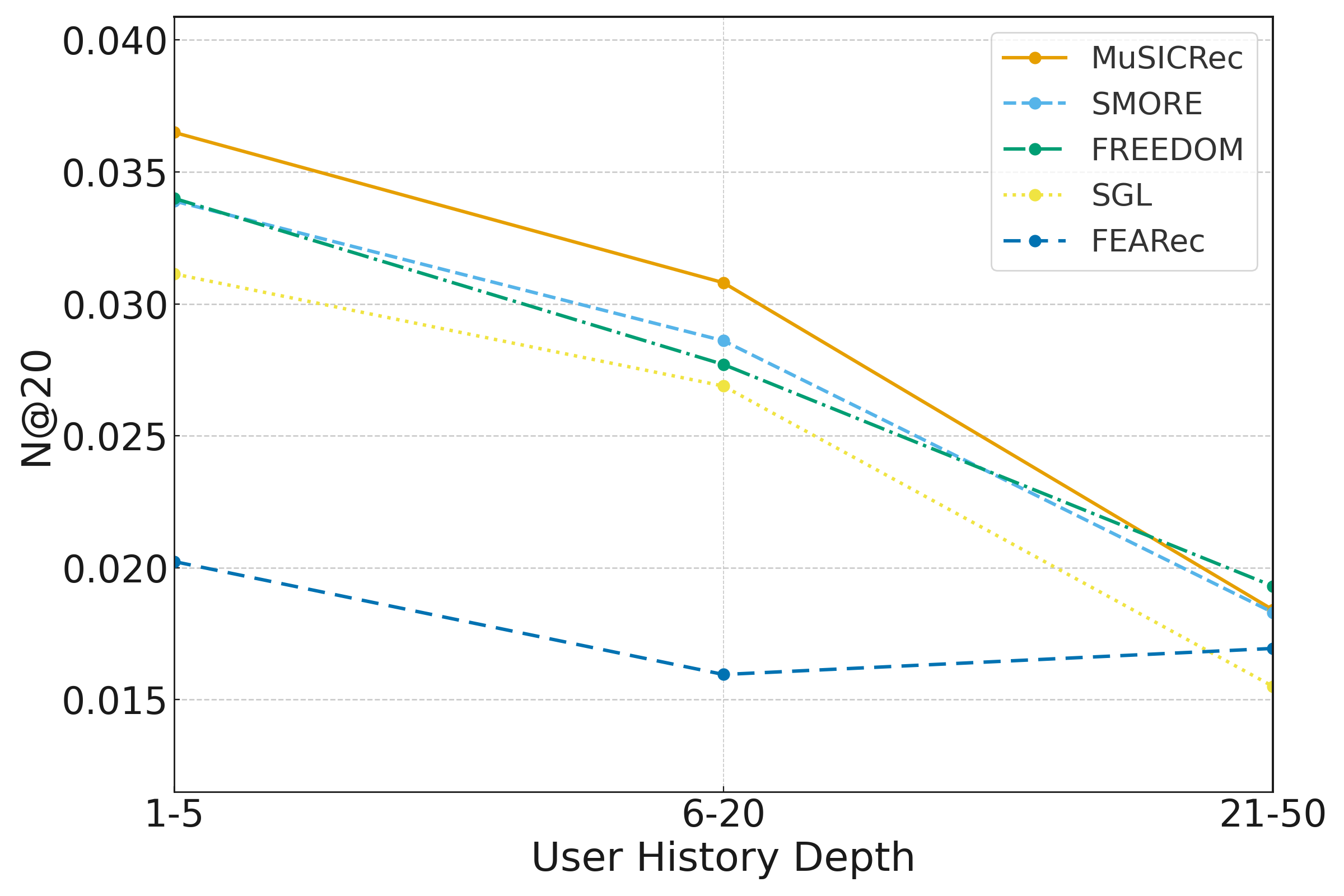}}\hfill
\subcaptionbox{\emph{Baby} R@20}{\includegraphics[width=0.32\textwidth]{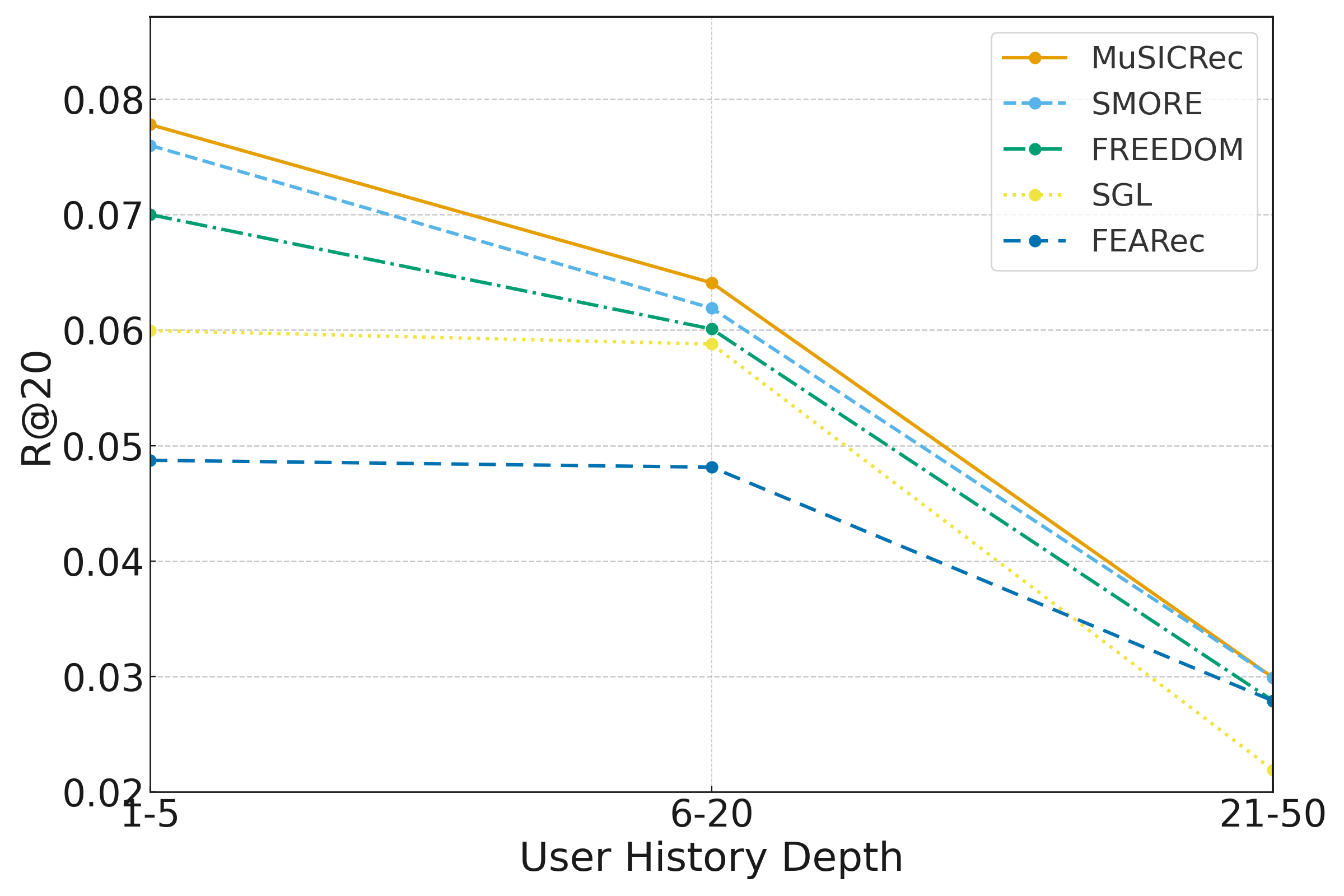}}\hfill
\subcaptionbox{\emph{Baby} N@20}{\includegraphics[width=0.32\textwidth]{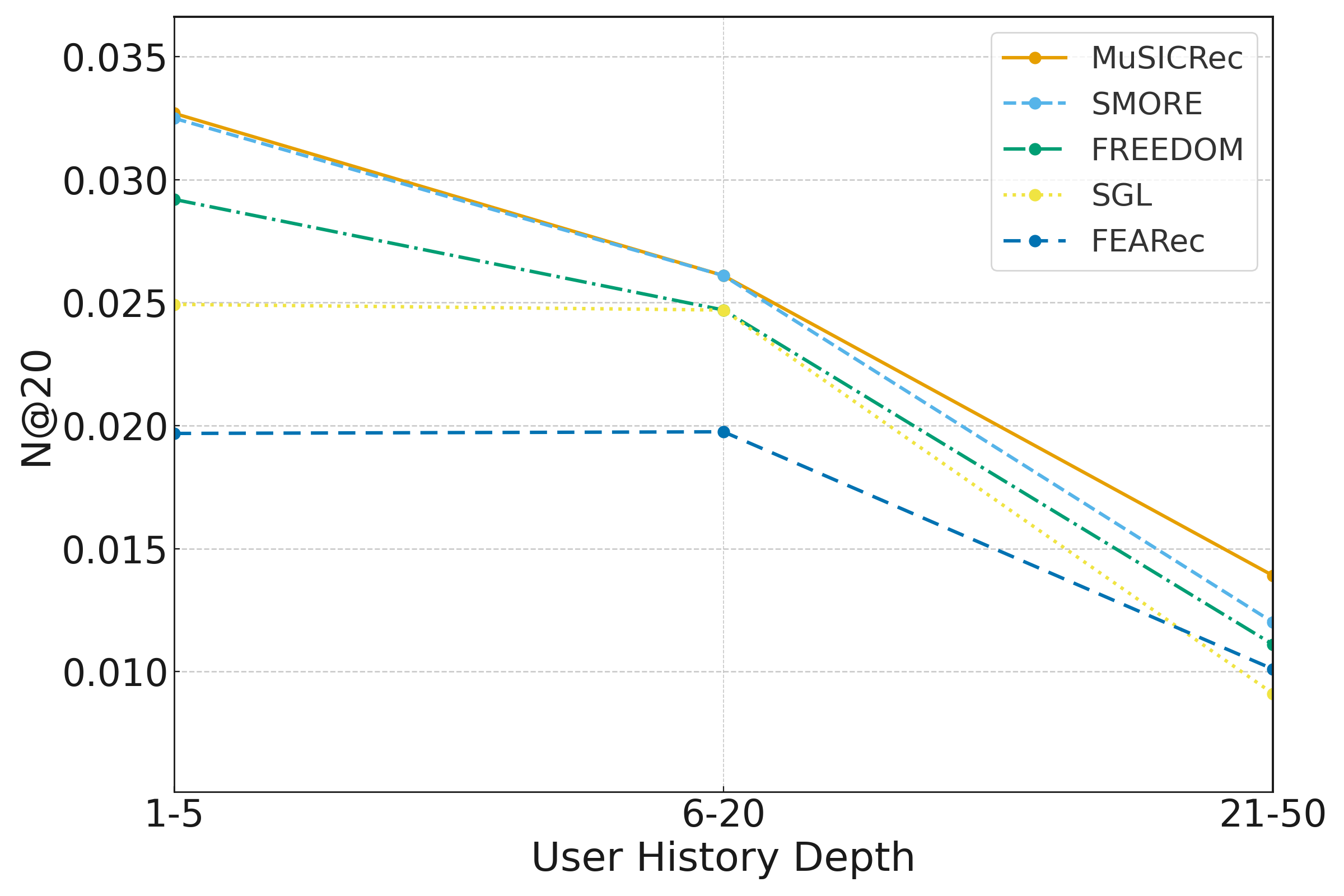}}\\

\subcaptionbox{\emph{Elec} R@20}{\includegraphics[width=0.32\textwidth]{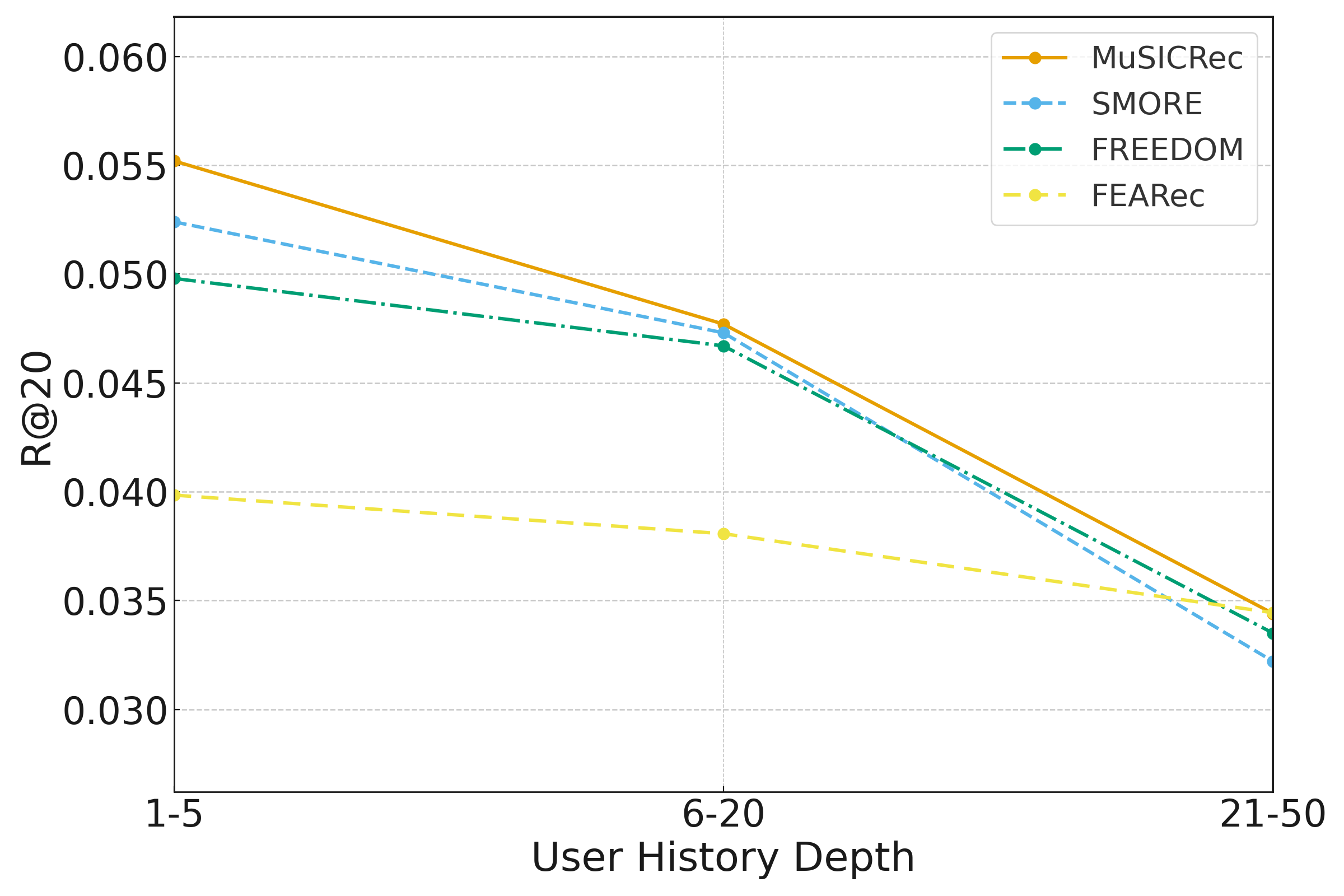}}\hfill
\subcaptionbox{\emph{Elec} N@20}{\includegraphics[width=0.32\textwidth]{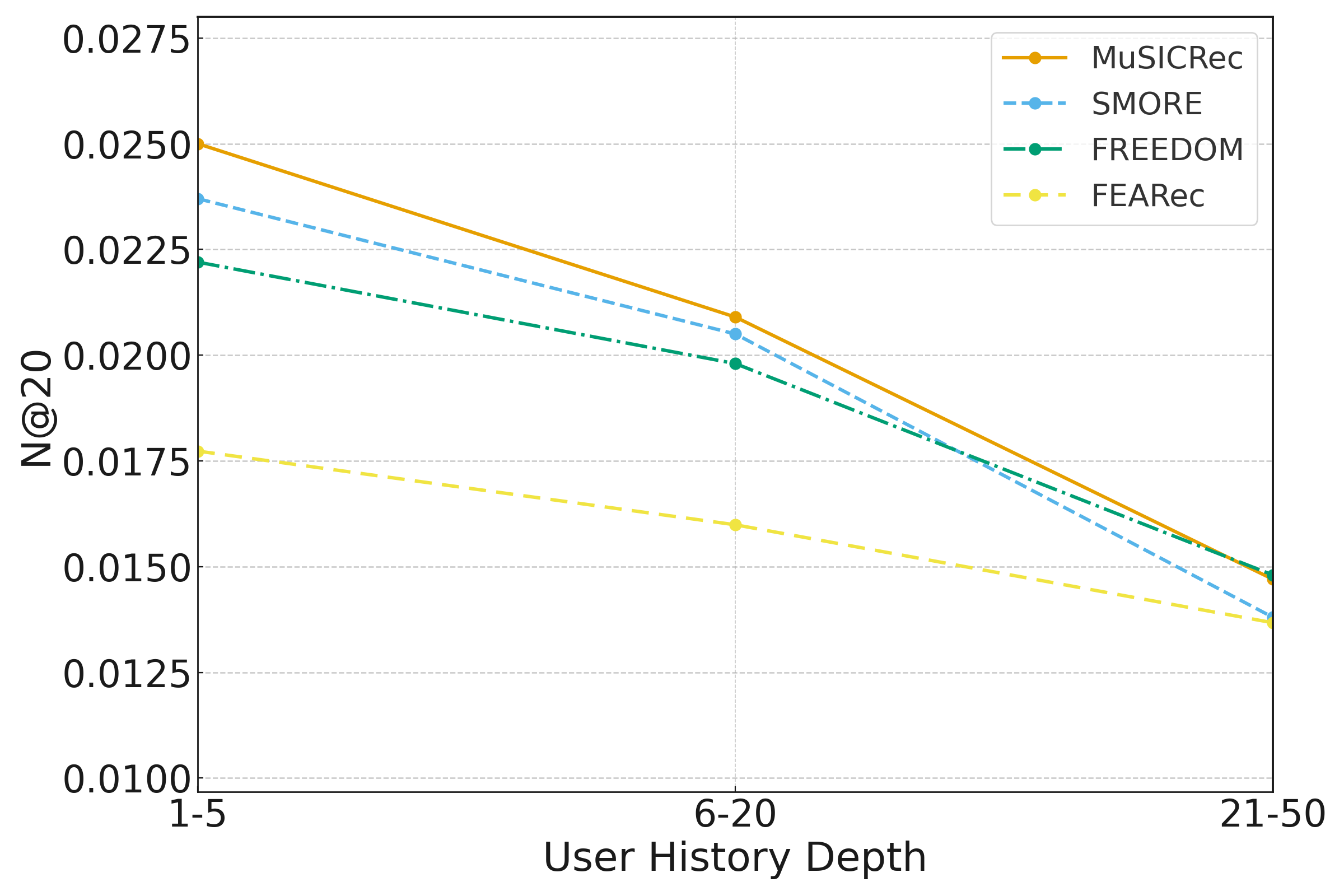}}\\

\captionof{figure}{User history depth analyses (additional results).}
\label{fig:app_userbucket}
]

\end{document}